\documentclass{IEEEtran}
\usepackage{amsmath,amsfonts}
\usepackage{verbatim}
\usepackage{graphicx}
\usepackage{subcaption}
\usepackage{hyperref}
\usepackage{multirow}
\usepackage[numbers,sort&compress]{natbib}
\usepackage[linesnumbered,ruled,vlined]{algorithm2e}
\usepackage{algorithmic}
\usepackage{float}
\usepackage[T1]{fontenc}
\usepackage{eso-pic}

\AtBeginDocument{%
  \addtolength\abovedisplayskip{-0.5\baselineskip}%
  \addtolength\belowdisplayskip{-0.5\baselineskip}%
}

\floatstyle{plaintop}
\restylefloat{table}

\SetCommentSty{mycommfont}

\AddToShipoutPictureBG{%
  \AtPageUpperLeft{%
    \put(0,-20){%
      \parbox{\paperwidth}{%
        \centering
        \textcolor{gray!100}{\fontsize{8}{12}\selectfont This work has 
        been published in the IEEE Access with DOI: 10.1109/ACCESS.2025.3586594.}\\
      }
    }
  }
}

\begin{document}
% \linenumbers

\title{GPS-Aided Deep Learning for Beam Prediction and Tracking in UAV mmWave Communication}

\author{Vendi Ardianto Nugroho and Byung Moo Lee,~\IEEEmembership{Senior Member,~IEEE}% <-this % stops a space
\thanks{The authors are with the Department of Intelligent Mechatronics Engineering, and Department of Convergence Engineering for Intelligent Drone, Sejong University, Seoul, 05006, South Korea, (e-mail: vendi@sju.ac.kr; blee@sejong.ac.kr).}
}

% Headers remain unchanged
\markboth{} %,~Vol.~x, No.~x, January~2025}%
{Shell \MakeLowercase{\textit{et al.}}: A Sample Article Using IEEEtran.cls for IEEE Journals}

\maketitle

\begin{abstract}
Millimeter-wave (mmWave) communication enables high data rates for
cellular-connected Uncrewed Aerial Vehicles (UAVs). However, a robust beam
management remains challenging due to significant path loss
and the dynamic mobility of UAVs, which can
destabilize the UAV-base station (BS) link. This research 
presents a GPS-aided deep learning (DL) model that 
simultaneously predicts current and future 
optimal beams for UAV mmWave communications, 
maintaining a Top-1 prediction accuracy exceeding 
70\% and an average power loss below 0.6 dB 
across all prediction steps.
These outcomes stem from a proposed data set splitting 
method ensuring balanced label distribution, 
paired with a GPS preprocessing technique that 
extracts key positional features, and a DL 
architecture that maps sequential position 
data to beam index predictions.
The model reduces overhead by approximately 93\% 
(requiring the training of 2 $\sim$
3 beams instead of 32 beams) with 95\% 
beam prediction accuracy guarantees,
and ensures 94\% to 96\% of predictions 
exhibit mean power loss not exceeding 1 dB.
\end{abstract}


\section{Introduction}
Uncrewed Aerial Vehicles (UAV) are expected to serve two roles in
wireless networks both as user equipment (UE) that
accesses cellular network (cellular-connected UAV) and as UAV-assisted communication
platforms providing aerial base stations (BS) and relays for
terrestrial users \cite{zeng-2019} \footnote{This work has been published in \cite{nugroho-2025}}. Enabling millimeter wave (mmWave) in
cellular-connected UAVs could support ultra high data traffic
and various UAV applications such as aerial surveillance
and disaster rescue due to the broad bandwidth in
the mmWave frequency band \cite{xiao-2022}. As the high path
loss characteristic of mmWave, deploying large antenna arrays
on the BS side helps mitigate it
by generating narrow beams with strong beamforming gains
\cite{geraci-2022}. As a result, mmWave communications rely heavily on efficient beam management—including beam training and tracking—to
quickly select the appropriate beams during intra- and
inter-cell mobility, minimizing the risk of beam misalignment
\cite{xue-2024b}. Moreover, UAV's high mobility introduces uncertainty in
parameters like velocity and acceleration, necessitating the development
of dedicated beam management methods tailored for UAV
systems \cite{zhang-2019a}. This highlights the need for innovative
approaches to overcome this challenge and enable high-speed
mmWave communication for UAV-assisted cellular networks. 

In mmWave communication systems, maintaining user connectivity relies on
two key steps: initial access and beam tracking \cite{giordani-2019},
\cite{yi-2024}. The initial access phase establishes the first
connection between BS and UE through beam training algorithms.
Once the initial beam direction is set, beam tracking
algorithms take over to maintain directional communication quality
and accelerate subsequent beam adjustments as the user moves.
The key challenges in beam management include high overhead
during the beam training process and maintaining beam alignment
under UE mobility, which can significantly impact system performance.
To tackle these challenges, researchers have proposed numerous 
approaches aimed at minimizing training overhead 
and improving the effectiveness of beam tracking systems.

Focusing first on the initial access phase, researchers have developed
multiple approaches to reduce beam training overhead through
various techniques
\cite{chen-2019,qi-2020,alkhateeb-2014,han-2016,alkhateeb-2018b,liu-2024a}. One method
involves hierarchical beam searching, as demonstrated by \cite{chen-2019}
and \cite{qi-2020}, which employs multi-layer codebooks where higher
layers utilize increasingly narrow beams, thus eliminating the
need for comprehensive narrow-beam combination searches. Another approach
focuses on compressive channel estimation \cite{alkhateeb-2014,han-2016}, which
aims to identify the strongest multi-path components Angle
of Arrival (AoA) and Angle of Departure (AoD)
pairs. Alternative solutions have emerged through deep learning
(DL) applications, including the approach of using pilot signals
received by multiple BS in \cite{alkhateeb-2018b}, and
method of combining instantaneous mmWave wide beam signals
with sub-6 GHz Channel State Information (CSI) for optimal beam estimation
in \cite{liu-2024a}. While these beam training approaches provide
initial beam alignment, maintaining optimal beam pairs during
user mobility requires effective beam tracking solutions. 

Various
beam tracking solutions have been proposed in the
literature \cite{xin-2019,larew-2019,ke-2019,yuan-2020,liu-2021c}. The study in
\cite{xin-2019} developed an Extended Kalman Filter (EKF) solution,
while research in \cite{larew-2019} implemented an Unscented Kalman
Filter (UKF) to sustain communication links. Taking a
different approach, research in \cite{ke-2019} introduced a Gaussian
process-based machine learning framework that enables swift and
precise UAV position prediction with measurable uncertainty levels.
This prediction helps constrain beam-tracking to specific spatial
areas around the predicted UAV positions. The study in
\cite{yuan-2020} focused on predicting angles between UAV-mounted BS and
UE, enabling both components to configure their transmit
and receive beams beforehand, thus enhancing communication reliability.
Whereas the study in \cite{liu-2021c} concentrated on UAV location
prediction, using the forecasted positions to determine optimal
angles between UAV and BS, facilitating efficient
beam alignment for subsequent time slots. 

Building upon
these advances in beam training and tracking, researchers
explored sensing-aided DL approaches that leverage sensor
data to further improve performance \cite{morais-2023,jiang-2023,
charan-2022a,luo-2023a}. The authors in \cite{morais-2023} 
utilized Global Positioning System
(GPS) data and applied a Multi Layer Perceptron
(MLP) model, while the study in \cite{charan-2022a} combines GPS
and vision data to perform current beam prediction. A Recurrent Neural Network
(RNN) based model was employed using LiDAR \cite{jiang-2023},
vision \cite{jiang-2022b}, and radar \cite{luo-2023a} to facilitate beam
tracking. While these approaches are effective to predict several
future beams, these works are designed for vehicle-to-infrastructure
(V2I) and vehicle-to-vehicle (V2V) scenarios which have different
movement characteristics compared to UAV. 

Recent research has
focused on the challenges of beam management in
UAV communications \cite{charan-2022,charan-2023,ahmad-2023,
zarei-2023,charan-2024}. The research in
\cite{charan-2022} applied a Convolutional Neural Network (CNN) and
MLP models using GPS and image modalities for
beam prediction, while studies in \cite{charan-2023,ahmad-2023,
zarei-2023} focused on image data with DL models
for the same purpose. However, due to UAV's
rapid movements, current beam predictions may become outdated
during data processing, making future beam prediction essential
for stable connectivity. The research presented in \cite{charan-2024}
introduced DL models for beam prediction and tracking,
utilizing GPS and visual data for UAV communications.
However, this approach needs separate models for the current
and future beam predictions which may increases computational
overhead. Additionally, while GPS-based prediction is crucial when
visual sensing fails (e.g., in low light), its
performance lags behind vision-based and beam vector-based DL
models. These challenges motivated our development of a
GPS-based DL model capable of simultaneous
current and future beam prediction for UAV communications.

This paper presents a lightweight GPS-based approach for mmWave beam prediction and 
tracking. The key contributions of this work include:

\begin{itemize}
    \item Designed a GPS-aided DL model architecture 
    that simultaneously performs beam prediction and tracking 
    in UAV mmWave communication scenario, leveraging 
    real-world GPS data from DeepSense6G data set \cite{alkhateeb-2023}.
    \item Proposed a data set splitting strategy called 
    \textit{adjusted splitting} to ensure similar data distribution 
    in training, validation, and test sets.
    \item Proposed a GPS data preprocessing method that combine 
    normalized UE's geodetic position with 
    UE-BS unit vector to obtain 
    optimal beam prediction and tracking performance.
\end{itemize}

The rest of this paper is organized as 
follows. Section II presents the system model
and the proposed solution. In Section III, the 
evaluation setup is introduced. In Section IV, 
the evaluation result is discussed. 
Finally, Section V concludes the paper.

\textit{Notations:} In this paper, 
boldface lower-case letters denote vectors.
For a vector $\mathbf{h}$, 
its Hermitian transpose is 
denoted as $\mathbf{h}^H$. 
$\mathbb{C}^{Q \times 1}$ denotes 
the space of $Q \times 1$ complex-valued vectors.
$\mathbb{E}[\cdot]$ 
denotes the expectation operator, 
and $\mathbb{P}\{\cdot\}$ represents probability. 
For a position vector $\mathbf{g}$, $\phi$ and 
$\lambda$ denote latitude and longitude respectively.
For time series data, $[t]$ denotes the value at 
time step $t$, and a sequence of $W$ consecutive 
observations is denoted as $\{[t-W+1], \dots, [t]\}$, 
where $W$ represents the observation window size. 
Predictions with step $v$ are denoted as $[t+v]$, 
where $v \in \{0,\dots,V\}$ and $V$ is the maximum 
prediction step.
$\mathcal{F} = \{\mathbf{f}_m\}_{m=1}^{M}$ denotes 
the beam codebook with $M$ beam vectors.
$\mathbf{u}_{UE-BS}$ represents the unit vector 
between UE and BS positions.

\section{System Model and Proposed Solution}
This section commences by presenting the underlying system
model that forms the foundation of our research.
We then outline the specific beam management tasks we address including GPS-aided current beam prediction and future beam prediction. We then proceed to
present our proposed solution.

\subsection{System Model}
\begin{figure}[t]
    \centering
    \includegraphics[width=0.5\textwidth]{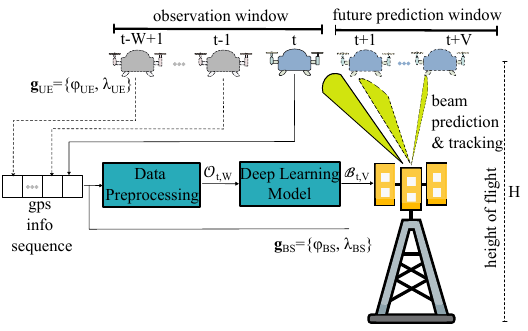}
    \caption{System Architecture}
    \label{fig:drone-schema}
\end{figure}

The system architecture, depicted
in Figure \ref{fig:drone-schema}, focuses on
the communication link between a UAV-mounted single antenna
(UE) and a BS equipped with
$Q$ antennas. The BS employs beamforming using vectors
selected from its codebook $\mathcal{F} = \{\mathbf{f}_m\}_{m=1}^{M}$, where
each $\mathbf{f}_m \in \mathbb{C}^{Q \times 1}$ and $M$
is the total number of beamforming vectors. This formulation includes beamforming 
at the BS, considering Time Division Duplexing (TDD) to facilitate the 
use of same frequency bands for both uplink and downlink transmissions. For an illustration,
at time step $t$, when the BS transmits a complex symbol 
$x[t] \in \mathbb{C}$ using a beam vector $\mathbf{f}[t]$, 
the received symbol at the UE can be expressed as:

\begin{equation}
y[t] = \mathbf{h}^H[t] \mathbf{f}[t]x[t] + n[t].
\label{eq:received-signal}
\end{equation}

In this expression, $\mathbf{h}[t] \in
\mathbb{C}^{Q \times 1}$ denotes the complex wireless channel
between the BS and UE antennas at time
$t$. The
downlink signal $x[t]$ complies with power constraint 
$\mathbb{E} \left[ x[t]^Hx[t] \right] = P$, where $P$ denotes
the transmit power. The term $n[t]$ is the receive noise which 
satisfies \( \mathbb{E} [n[t] n^H [t]] = \sigma^2_n \), 
where \( \sigma^2_n \) represents the receive noise power.

\subsection{Problem Formulation and DL Task Definitions}

This study adopts the problem formulation and DL
task definitions from \cite{jiang-2022b} and \cite{jiang-2023} to define
the optimal beam vector and index, as well
as the beam prediction and beam tracking tasks. The BS achieves optimal 
beam selection by choosing the beam vector $\mathbf{f}$ 
that generates the highest received power.
The ideal beam vector $\mathbf{f^*}[t]$ is expressed as: 
\begin{equation}
\mathbf{f^*}[t] =
\underset{ \mathbf{f}[t] \in \mathcal{F} }{ \operatorname{arg\,max\,} } |\mathbf{h}^H[t]
\mathbf{f}[t]|^2. 
\label{eq:beam-vector}
\end{equation}
Under this
pre-defined codebook constraint, the optimal beam vector $\textbf{f}^*[t]$
can be uniquely identified by its corresponding beam
index within the codebook. At time step $t$,
the optimal beam index $b^{*}[t]$ is determined by
the following condition: 
\begin{equation}
b^*[t] = \underset{ b[t] \in \{1,2,\dots,M\} }{ \operatorname{arg\,max\,} } |\mathbf{h}^H[t] \mathbf{f}_{b[t]}|^2,
\label{eq:beam-index}
\end{equation}
where $M$ represents the size of codebook $\mathcal{F}$ and $\mathbf{f}_{b[t]}$ represents the $b[t]$-th beam vector from the codebook $\mathcal{F}$ at time $t$.
It is worth emphasizing that under the codebook
constraint, determining the optimal beam is equivalent to
identifying its corresponding optimal beam index. 

This study
investigates utilizing UAV positional information to enable beam
prediction (estimating current beams) and beam tracking (forecasting
future beams). The two-dimensional position vector at time $t$,
represented as $\mathbf{g}_{UE,norm}[t] \in \mathbb{R}^2$, consists of UE's normalized
latitude $\phi_{UE,norm}[t]$ and longitude $\lambda_{UE,norm}[t]$. Additionally, $\mathbf{u}_{UE-BS}[t] \in
\mathbb{R}^3$ is defined as a unit vector derived
from the positions of the UE and BS
at time $t$. Further details on the definitions
of $\mathbf{g}_{UE,norm}[t]$ and $\mathbf{u}_{UE-BS}[t]$ are provided in subsection
\ref{sec:gps-data-preprocessing}.

To predict both current and future beams, these
representations are combined as $\mathbf{O}[t] = \{\mathbf{g}_{UE,norm}[t], \mathbf{u}_{UE-BS}[t]\}$.
The sequence of observed sensory data within an
observation window $W \in \mathbb{Z^+}$ is defined as
$\mathcal{O}_{t,W} = \{\mathbf{O}[t-W+1], \dots, \mathbf{O}[t]\}$. Using the definition
of the optimal beam index provided in equation
\begin{equation}
    \underset{\hat{\mathcal{B}}_{t,V}}{\operatorname{arg\,max\,}} \mathbb{P}\{\hat{\mathcal{B}}_{t,V} = \mathcal{B}^*_{t,V} | \mathcal{\mathcal{O}}_{t,W}\}.
\label{eq:beam-pred-problem}
\end{equation}

In this context, $\mathbb{P}\{\cdot|\cdot\}$
denotes the conditional probability. The term $\hat{\mathcal{B}}_{t,V}$ represents
the predicted beam index $\hat{b}[t]$ sequence starting from time step $t$ to $t+V$, the term $\mathcal{B}^*_{t,V}$ represents
the optimal beam index $b^*[t]$ sequence starting from time step $t$ to $t+V$, while
$\mathcal{O}_{t,W}$ refers to the observed sensory data from
time step $t-W+1$ through $t$. Furthermore, $v \in \{0, 1, 2, \dots, V\}$ represents 
the prediction step in the beam prediction and tracking process. Below, we provide 
a formal definition of the DL task for beam prediction and tracking.

\textbf{Beam Prediction and Tracking DL Task Definition}: In this approach,
we use a DL model to solve equation
\eqref{eq:beam-pred-problem}. To address the beam prediction problem, the
goal of the DL model is to predict
the optimal beam index sequence $\mathcal{B}^*_{t,V}$ based on the
sensory data sequence $\mathcal{O}_{t,W}$. Consequently, the optimal DL model
for GPS-aided beam prediction and tracking can be expressed as:
\begin{equation}
    f_{bpt}^{*}(\mathcal{O}_{t,W};\Theta_{bpt}^{*}) = \underset{f_{bpt}(;\Theta_{bpt})}{\operatorname{arg\,max\,}} \mathbb{P}\{f_{bpt}(\mathcal{O}_{t,W};\Theta_{bpt}) = \mathcal{B}^*_{t,V}\},
\label{eq:beam-pred-track-opt}
\end{equation}
where $f^{*}_{bpt}(\cdot)$ represents the
DL optimal model's mapping function for the beam prediction
and tracking task, and $\Theta^{*}_{bpt}$ refers to the optimal model's training
parameters. 

\subsection{Proposed Solution}
In this subsection, we introduce how the GPS
data being processed before used as the DL
model input and the architecture of DL model. 

\begin{figure}[t]
    \centering
    \includegraphics[width=\columnwidth]{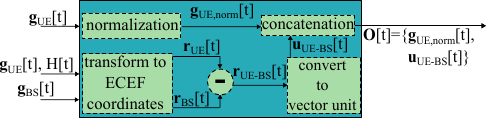}
    \caption{GPS data preprocessing pipeline for the proposed beam prediction and tracking system}
    \label{fig:data-preprocessing}
\end{figure}

\subsubsection{GPS Data Preprocessing}
\label{sec:gps-data-preprocessing}

Given that raw GPS sequence data 
implicitly encodes spatiotemporal information, 
preprocessing is essential before inputting 
the data into a DL model to 
achieve optimal performance \cite{dong-2016}. The block 
diagram in Figure \ref{fig:data-preprocessing}
block illustrates our approach to preprocess GPS data
collected from UE and BS. Initially, we normalize
the UE's position using min-max normalization. Subsequently, we
transform geodetic coordinates to Earth-centered Earth-fixed (ECEF) coordinates
and extract the UE-BS vector unit. The final
step involves concatenating these features. In the following
sections, we elaborate on the min-max normalization process
and the technique for obtaining ECEF coordinates.

\begin{enumerate}[a)]
    \item Min-max Normalization: 
    Consider a data set $\mathcal{D}$ comprising $K$ sample pairs
    of UE's latitude and longitude 
    $\mathbf{g}_{UE,k}=(\phi_{UE,k},\lambda_{UE,k})^{K}_{k=1}$.
    We identify $\phi_{UE,min}$ and $\phi_{UE,max}$ as the minimum and maximum
    latitude values in data set $\mathcal{D}$, and $\lambda_{UE,min}$ and
    $\lambda_{UE,max}$ as the minimum and maximum longitude values.
    The normalization process transforms each UE's latitude and
    longitude to a value between 0 and 1
    using the following equations: 
    \begin{equation}
    \phi_{UE,norm,k} = \frac{\phi_{UE,k} - \phi_{UE,min}}{\phi_{UE,max}- \phi_{UE,min}},
    \label{eq:minmax-lat}
    \end{equation}
    \begin{equation}
    \lambda_{UE,norm,k} = \frac{\lambda_{UE,k} - \lambda_{UE,min}}{\lambda_{UE,max}- \lambda_{UE,min}},
    \label{eq:minmax-lon}
    \end{equation}
    
    \item ECEF Coordinate Transformation: 
    Drawing from Paul's work in \cite{paul-1973} as referenced in \cite{zhu-1994},
    the ECEF coordinates $(\alpha,\beta,\gamma)$ of a point $\mathbf{r}$ can be
    computed from its geodetic coordinates ($\phi, \lambda,A$) using
    the following mathematical transformations: 
    \begin{equation}
    \alpha = (R+A) \cos \phi \cos \lambda,
    \label{eq:alpha}
    \end{equation}
    \begin{equation}
    \beta = (R+A) \cos \phi \sin \lambda,
    \label{eq:beta}
    \end{equation}
    \begin{equation}
    \gamma = (R+A-e^2R) \sin \phi,
    \label{eq:gamma}
    \end{equation}
    where
    \begin{equation}
    R = \frac{a}{\sqrt{1-e^2 \sin^2 \phi}}.
    \label{eq:R}
    \end{equation}
    The parameters in these equations represent: $\phi$ 
    as geodetic latitude, $\lambda$ as geodetic longitude, 
    $A$ as the altitude normal to the ellipsoid, $a$ as 
    the ellipsoidal equatorial radius ($a=6378.137$ km for WGS-84 model), 
    and $e$ as the ellipsoid's eccentricity ($e^2=0.00669437999$ for 
    WGS-84 model). We define
    $\mathbf{r}_{BS}[t]$ and $\mathbf{r}_{UE}[t]$ as the ECEF coordinates of
    BS and UE at time $t$ respectively. The
    UE-BS vector $\mathbf{r}_{UE-BS}[t]$ is expressed as: 
    \begin{equation}
    \mathbf{r}_{UE-BS}[t] = \mathbf{r}_{UE}[t] - \mathbf{r}_{BS}[t]. 
    \label{eq:rvec}
    \end{equation}
    With the magnitude of $\mathbf{r}[t]$ is calculated as
    $|\mathbf{r}[t]|=\sqrt{ \alpha^2+\beta^2+\gamma^2 }$, we express the unit vector
    of the UE-BS vector $\mathbf{u}_{UE-BS}[t]$ as: 
    \begin{equation}
    \mathbf{u}_{UE-BS}[t] = \frac{\mathbf{r}_{UE-BS}[t]}{|\mathbf{r}_{UE-BS}[t]|}. 
    \label{eq:rvec-unit}
    \end{equation}
    
\end{enumerate}

\begin{figure}[t]
    \centering
    \includegraphics[width=0.7\columnwidth]{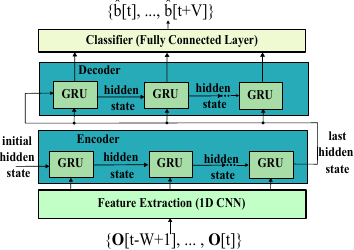}
    \caption{Neural network architecture used for beam prediction and tracking}
    \label{fig:drone-model}
\end{figure}

\subsubsection{DL Model Architecture}
As illustrated in Figure \ref{fig:drone-model}, the proposed DL 
model architecture consists of three primary components: 
the Feature Extraction Block, the Encoder-Decoder RNN Block, 
and the Classifier Block. Since 1D CNNs excel at extracting 
local features 
from sequential data \cite{ige-2024}, the feature extraction 
block utilizes 1D CNN layers to process GPS sequence data $\mathcal{O}_{t,W}$, 
facilitating the detection of UAV movement patterns.
The kernel size is configured to 3 to capture local features 
over each 3 of time steps, while the kernel advances 
with a stride of 1 for ensuring detailed coverage. To 
maintain the original sequence length post-feature 
extraction, we add zero padding at both the start
and end of the input data. Table~\ref{tab:model-params} 
provides a comprehensive overview 
of the network architecture and its parameters. 
It specifies the input and output feature 
channels ($c_i$ and $c_o$ respectively), alongside 
the kernel size, stride, and padding parameters 
($k_1$, $k_2$, $k_3$). The architecture incorporates 
batch normalization (BN) layers to standardize the 
data distribution to zero mean and unit variance.

The encoder-decoder RNN architecture
uses two RNNs: an encoder that compresses a variable-length 
input sequence into a fixed-dimensional context vector, 
and a decoder that generates a variable-length output 
sequence from this vector \cite{cho-2014}. Since this architecture can 
effectively transform sequences (e.g., for language translation), 
we adapt it to predict beam index sequence from GPS sequence. 
Our encoder processes a $W$-step GPS sequence, capturing 
temporal dependencies through hidden state updates.
The encoder's final hidden state serves as the context vector 
for the decoder. This context vector is replicated $V+1$ times
to serve as the decoder's input sequence.
Since GRUs offer superior 
convergence properties over standard RNNs as demonstrated 
in \cite{chung-2014}, 
we implement GRU in both the encoder and decoder.

The classifier block consists of two fully connected (FC)
layers with a softmax activation function to predict current 
optimal beam index and 
future beam index $ \hat{\mathcal{B}}_{t,V} = \{\hat{b}[t], \dots, \hat{b}[t+V]\}$. This approach
generates a score vector $\mathbf{\hat{c}} = [c_{1}, \dots
, c_{M}]^T$, where each score $c_{m} \in (0,
1)$ corresponds to a specific $m$-th beam $b_{m}$ in
the codebook. The optimal future beam is determined
by selecting the beam index with the highest
score

\begin{equation}
    \hat{b} = \underset{b \in \{1, \dots, M\}}{\operatorname{arg\,max\,}}\; \mathbf{c}_b.
    \label{eq:optimal-beam-index}
\end{equation}

\textbf{Learning Process}: The DL model undergoes offline 
supervised training, with
each data point structured in two key components
: (i) an input sequence of position data $\{\mathbf{O}[t-W+1],
\dots, \mathbf{O}[t]\}$ , and (ii) a label/desired output
sequence $\{\mathbf{c}^*[t], \dots, \mathbf{c}^*[t+V]\}$ where each $\mathbf{c}^*[t+v]$ is
a one-hot representation of $b^*[t+v]$. To address the
classification problem, a cross-entropy loss function 
is implemented. For each sample in the data set, 
the loss function is mathematically expressed as

\begin{equation}
\mathcal{L} = - \sum_{j=t}^{t+V} \sum_{m=1}^{M} b_m^{*}[j] \log_e(\hat{c}_m[j]),
\label{eq:ce-loss}
\end{equation}
where $b^*_{m}[j]$ denotes the $m$-th element of the
one-hot coded vector $\mathbf{b}^*$ at time step $j$,
and $\hat{c}_{m}[j]$ represents the $m$-th element of the
output vector $\hat{\mathbf{c}}[j]$ at the same time step. The loss for 
each sample is subsequently averaged across the entire data set.
The training is configured with specific hyperparameters: 20
epochs, a train batch size of 8, a validation batch size of 1024, a test batch size of
1024, an Adam optimizer with a learning rate of
$5 \times 10^{-4}$, a zero weight decay, and a
learning rate reduction factor of 0.1 on epoch
12 and 18.

\begin{table}[t]
    \caption{Structure and parameters of the proposed network}
    \label{tab:model-params}
    \resizebox{\columnwidth}{!}{
        \begin{tabular}{lll}
            \hline
            Module                              & Layer              & Parameter                         \\ \hline
            \multirow{3}{*}{Feature Extraction} & 1D Convolution     & $c_i$=5, $c_o$=128, (3,1,1)       \\
                                                & 1D Convolution     & $c_i$=128, $c_o$=128, (3,1,1), BN \\
                                                & ReLU Activation    & $c_i$=128, $c_o$=128              \\ \hline
            Encoder                             & GRU                & $c_i$=128, $c_o$=128              \\ \hline
            Decoder                             & GRU                & $c_i$=128, $c_o$=128              \\ \hline
            \multirow{2}{*}{Classifier}         & FC                 & $c_i$=128, $c_o$=64               \\
                                                & ReLU Activation    & $c_i$=64, $c_o$=64                \\
                                                & FC                 & $c_i$=64, $c_o$=32               \\
                                                & Softmax Activation & $c_i$=32, $c_o$=32                \\ \hline
        \end{tabular}
    }
\end{table}

\section{Evaluation Setup}
This study leverages Scenario 23 from the DeepSense
6G data set \cite{alkhateeb-2023} to investigate high-frequency
wireless communication applications involving drones. The BS is
configured with a 16-element phased array operating in
the 60 GHz frequency band, utilizing a codebook
comprising 32 pre-defined beams ($M=32$). The UE takes
the form of a remotely controlled drone equipped
with a mmWave transmitter, a GPS receiver, and
inertial measurement units (IMU). The transmitter features a
quasi-omnidirectional antenna that continuously broadcasts signals omnidirectionally at
the 60 GHz frequency band.
Note that this study assumes uplink/downlink
reciprocity in received power, 
as mmWave signals typically follow a dominant single path
carrying most of the power. In the subsequent
sections, we will elaborate on the data splitting
methodology and data transformation techniques employed to prepare
the data set for our DL model.

\begin{figure*}[t]
    \centering
    \includegraphics[width=0.9\textwidth]{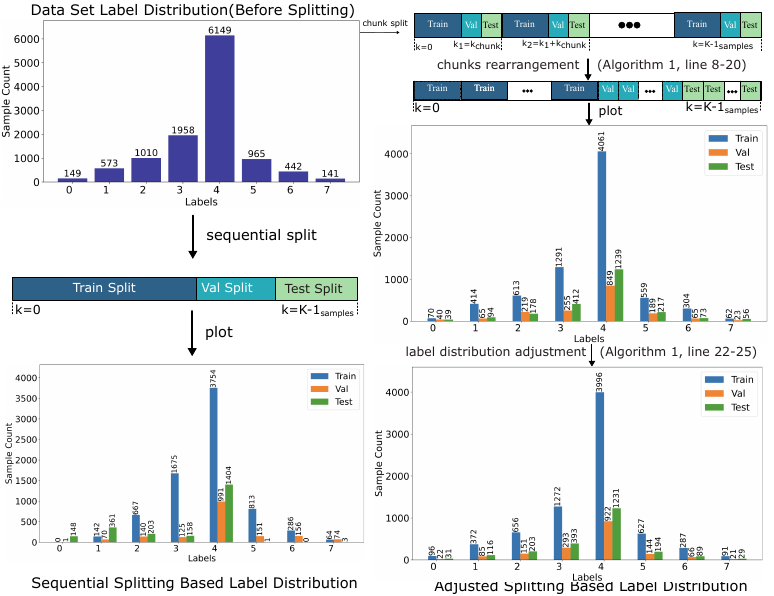}
    \caption{Impact of data set splitting methods on label distribution across train, validation, and test sets}
    \label{fig:dataset-split}
\end{figure*}

\subsection{Data Set Splitting Method}

We consider a raw data set $D_{raw}$ comprises $K$
samples $D_{raw} =\{(q_{k},t_{k},\phi_{BS,k},\lambda_{BS,k},\phi_{UE,k},
\lambda_{UE,k},H_{k},\\ \mathbf{u}_{UE-BS,k}, b^*_{k})\}^K_{k}$, 
which encapsulate various key
parameters. These parameters include the sequence index $q$,
sample index $t$, BS's geodetic coordinates ($\phi_{BS,k},\lambda_{BS,k}$), UE's
geodetic coordinates ($\phi_{UE,k},\lambda_{UE,k}$), the vertical distance between the
drone and ground $H_{k}$, the UE-BS unit vector
$\mathbf{u}_{UE-BS,k}$, and the actual beamforming index $b^*_{k}$. Note that the sequence index $q$ is a label 
that groups together all the data samples collected during a single trip of 
the UE as it moves in front of the BS \cite{alkhateeb-2023}.

Employing equations \eqref{eq:minmax-lat} and \eqref{eq:minmax-lon}, we normalize the UE's positioning. To calculate the UE-BS unit 
vector $\mathbf{u}_{UE-BS,k}$ with \eqref{eq:rvec-unit}, 
first we determine the UE's ECEF coordinates using \eqref{eq:rvec}, where 
the UE's altitude is set to $H_{k}$ and the BS's altitude to $0$. The
raw data set $D_{raw}$ is subsequently partitioned into three
distinct subsets: training ($D^{train}_{raw}$), validation ($D^{val}_{raw}$), and
testing ($D^{test}_{raw}$) data sets, allocated according to predetermined percentage
portions. Within each
split data set, samples are arranged in ascending order
based on their sample index $t_{k}$. Our objective
is to compare two different data splitting strategy
methods.

\textbf{\SetAlFnt{\small}
\SetAlCapFnt{\large}
\SetAlCapNameFnt{\large}
\begin{algorithm}
    \algsetup{linenosize=\tiny}
    \SetKwInOut{Input}{Input}
    \SetKwInOut{Output}{Output}
    \SetKwComment{Comment}{$\triangleright$\ }{}
    
    \underline{function AdjustedSplit}($\mathcal{D}_{\text{raw}}, f_{\text{train}}, f_{\text{val}}, f_{\text{test}}, P_{\text{chunks}}$)\;
    \Input{
        $\mathcal{D}_{\text{raw}}$: Raw data set\;
        $f_{\text{train}}, f_{\text{val}}, f_{\text{test}}$: Split ratios\;
        $P_{\text{chunks}}$: Chunk size percentages\;
        $M$: Beam codebook size\;
    }
    \Output{
        $\mathcal{D}^{\text{train}}_{\text{raw}}, \mathcal{D}^{\text{val}}_{\text{raw}}, \mathcal{D}^{\text{test}}_{\text{raw}}$
    }
    
    \BlankLine
    $N_{\text{total}} \leftarrow |\mathcal{D}_{\text{raw}}|$\;
    Get label distribution $P(b)$ from $\mathcal{D}_{\text{raw}}$\;
    Calculate chunk sizes $C_{\text{sizes}}$ from $P_{\text{chunks}}$\;
    Filter $C_{\text{sizes}}$ based on minimum sequence length\;
    $s_{\text{best}} \leftarrow \infty$\;
    $\mathcal{D}^{\text{train}}_{\text{best}}, \mathcal{D}^{\text{val}}_{\text{best}}, \mathcal{D}^{\text{test}}_{\text{best}} \leftarrow \emptyset$\;
    
    \ForEach{chunk size $c \in C_{\text{sizes}}$}{
        Initialize $\mathcal{D}^{\text{train}}_{\text{raw}}, \mathcal{D}^{\text{val}}_{\text{raw}}, \mathcal{D}^{\text{test}}_{\text{raw}} \leftarrow \emptyset$\;
        Split raw data into chunks of size $c$\;
        \ForEach{chunk}{
            Split chunk into $\mathcal{D}^{\text{train}}_{\text{chunk}}, \mathcal{D}^{\text{val}}_{\text{chunk}}, \mathcal{D}^{\text{test}}_{\text{chunk}}$ using given ratios\;
            $\mathcal{D}^{\text{train}}_{\text{raw}} \leftarrow \mathcal{D}^{\text{train}}_{\text{raw}} \cup \mathcal{D}^{\text{train}}_{\text{chunk}}$\;
            $\mathcal{D}^{\text{val}}_{\text{raw}} \leftarrow \mathcal{D}^{\text{val}}_{\text{raw}} \cup \mathcal{D}^{\text{val}}_{\text{chunk}}$\;
            $\mathcal{D}^{\text{test}}_{\text{raw}} \leftarrow \mathcal{D}^{\text{test}}_{\text{raw}} \cup \mathcal{D}^{\text{test}}_{\text{chunk}}$\;
        }
        Calculate label distributions $P_{\text{train}}(b), P_{\text{val}}(b), P_{\text{test}}(b)$ for $\mathcal{D}^{\text{train}}_{\text{raw}}, \mathcal{D}^{\text{val}}_{\text{raw}}, \mathcal{D}^{\text{test}}_{\text{raw}}$\;
        Get distribution similarity score  $s \leftarrow \sum_{m=1}^{M} |P(b) - P_{\text{train}}(b)| + |P(b) - P_{\text{val}}(b)| + |P(b) - P_{\text{test}}(b)|$\;
    
        \If{$s < s_{\text{best}}$}{
            $s_{\text{best}} \leftarrow s$\;
            $\mathcal{D}^{\text{train}}_{\text{best}} \leftarrow \mathcal{D}^{\text{train}}_{\text{raw}}$, $\mathcal{D}^{\text{val}}_{\text{best}} \leftarrow \mathcal{D}^{\text{val}}_{\text{raw}}$, $\mathcal{D}^{\text{test}}_{\text{best}} \leftarrow \mathcal{D}^{\text{test}}_{\text{raw}}$\;
        }
    }
    
    $\mathcal{D}^{\text{train}}_{\text{raw}} \leftarrow \mathcal{D}^{\text{train}}_{\text{best}}$, $\mathcal{D}^{\text{val}}_{\text{raw}} \leftarrow \mathcal{D}^{\text{val}}_{\text{best}}$, $\mathcal{D}^{\text{test}}_{\text{raw}} \leftarrow \mathcal{D}^{\text{test}}_{\text{best}}$\;
    
    \ForEach{label $b \in \{1, ..., M\}$}{
        $\mathcal{D}_b \leftarrow$ Get samples for label $b$ from $\mathcal{D}^{\text{train}}_{\text{raw}}, \mathcal{D}^{\text{val}}_{\text{raw}}, \mathcal{D}^{\text{test}}_{\text{raw}}$\;
        Split $\mathcal{D}_b$ into $\mathcal{D}^{\text{train}}_b, \mathcal{D}^{\text{val}}_b, \mathcal{D}^{\text{test}}_b$ using given ratios\;
        Update $\mathcal{D}^{\text{train}}_{\text{raw}}, \mathcal{D}^{\text{val}}_{\text{raw}}, \mathcal{D}^{\text{test}}_{\text{raw}}$ with new label-based splits\;
    }
    
    \Return{$\mathcal{D}^{\text{train}}_{\text{raw}}, \mathcal{D}^{\text{val}}_{\text{raw}}, \mathcal{D}^{\text{test}}_{\text{raw}}$}\;
    
    \caption{Adjusted Splitting}
    \label{algo:algo1}
\end{algorithm}}

\subsubsection{Sequential Splitting}
We employ a straightforward method that preserves the
original sample index order by sequentially partitioning $D_{raw}$.
Specifically, the first $65\%$ of $D_{raw}$ 
becomes $D^{train}_{raw}$,
the next $15\%$ becomes $D^{val}_{raw}$, and the remaining 
$20\%$ becomes $D^{test}_{raw}$.

\subsubsection{Adjusted Splitting}
In this approach, we split $D_{raw}$ by considering
the label distribution (optimal beam index $b^*$) through
Algorithm \ref{algo:algo1}. The primary goal is to ensure
that the label distribution of each split data set 
reflects the label distribution of the original data set $D_{raw}$ and
follows the desired splitting percentage. The methodology unfolds
through multiple stages: (i) We first divide the raw
data set $D_{raw}$ into multiple chunks. Each chunk is divided into train 
chunk data set $D^{train}_{chunk}$, validation chunk data set $D^{val}_{chunk}$,
and test chunk data set $D^{test}_{chunk}$, which are then aggregated to form $D^{train}_{raw}$,
$D^{val}_{raw}$, and $D^{test}_{raw}$. Recognizing that simple grouping might
not guarantee the desired label distribution, we introduce an
additional refinement. (ii) We organize split data sets into label-based
grouped data sets $D_{group} = \{D_{b}\}^{M}_{b}$, where $b$ represents
the label (optimal beam index). Within each $D_{b}$,
we implement a consistent $65\%$, $15\%$, and $20\%$
split into train $D^{train}_{b}$, validation $D^{val}_{b}$, and test $D^{test}_{b}$ data sets. We then consolidate all $D^{train}_{b}$ into
$D^{train}_{raw}$, $D^{val}_{b}$ into $D^{val}_{raw}$, and $D^{test}_{b}$ into $D^{test}_{raw}$.

The illustration in Figure \ref{fig:dataset-split} highlights the impact
of data set splitting methods on label distribution across
train, validation, and test sets. We examine the
label distribution on DeepSense 6G scenario 23 data set
using a beam codebook with size $M=8$ and applying a split ratio of $65\%$, $15\%$, and $20\%$
for training, validation, and testing, respectively. The sequential splitting method
exposes significant label distribution imbalances. 
Notably, the 0-th beam index presents an issue: the 
train data set is empty, and 
the test data set contains a lot of samples. In 
contrast, the adjusted splitting method ensures 
that the label distribution in each split 
data set reflects the label distribution of 
the original data set and follows the desired split ratio. 

\subsection{Data Transformation}
We transform each split data set into a development
data set containing $U$ samples $D_{dev}=\{\mathcal{O}_{q,t,W,u},\mathcal{B}^*_{q,t,V,u}\}^U_{u}$, which represent the
model's input $\mathcal{O}_{q,t,W,u}$ and its corresponding desired output
$\mathcal{B}^*_{q,t,V,u}$. The input $\mathcal{O}_{q,t,W,u}$ represents a sequence of
observed sensory data $\mathcal{O}_{q,t,W,u}= \{\mathbf{O}_{q,u}[t-W+1], \dots, \mathbf{O}_{q,u}[t]\}$, where
$W =8$ and $\mathbf{O}_{q,u}[t]= \{\mathbf{g}_{UE,norm}[t], \mathbf{u}_{UE-BS}[t]\}$. The output
$\mathcal{B}^*_{q,t,V,u}$ denotes the sequence of current and future
beam indices $\mathcal{B}^*_{q,t,V,u}= \{b^*_{q,u}[t], \dots, b^*_{q,u}[t+V]\}$, with $V = 3$. We implement constraints to ensure the
sequence integrity, requiring that the sequences $\mathcal{O}_{q,t,W,u}$ and
$\mathcal{B}^*_{q,t,V,u}$ must share the same sequence index $q$
and be ordered consecutively based on sample index
$t$. For instance, this could manifest as an
input sequence $\{\mathbf{O}_{q=1,u=1}[t-W+1], \dots, \mathbf{O}_{q=1,u=1}[t] \}$  with a corresponding
output $\{b^*_{q=1,u=1}[t], \dots, b^*_{q=1,u=1}[t+V]\}$. In cases where creating a perfect consecutive sequence is
not possible, we implement a technique to guarantee
at least $V+1$ consecutive samples with identical
sequence index $q$. This is achieved by padding the
initial $W-1$ elements of the input sequence with
a zero vector $\textbf{Z}$ of length $5$,
matching the dimensions of $\mathbf{O}$. For instance, this method might generate
an input sequence such as $\{\mathbf{Z}_{q=1,t=0,W=8,u=1}, \dots, \mathbf{Z}_{q=1,t=6,W=8,u=1}, \\ \mathbf{O}_{q=1,t=7,W=8,u=1}\}$,
while maintaining consistency in the output sequence. This strategy
ensures the model is provided with well-structured inputs even
when seamless continuity is unattainable. As not every sample
in the raw data set $D_{raw}$ can be arranged into these necessary sequences, the overall sample count
in the development data set $D_{dev}$ will be smaller than the total 
number contained in the raw data set $D_{raw}$ (see Table \ref{tab:dataset-detail}).

\begin{table}[t]
    \caption{Data Set Details}
    \centering
    \label{tab:dataset-detail}
    \resizebox{0.8\columnwidth}{!}{
    \begin{tabular}{||l||ll||}
    \hline \hline
    \multicolumn{1}{||c||}{\multirow{2}{*}{\textbf{\begin{tabular}[c]{@{}c@{}}Data Set\\ Type\end{tabular}}}} & \multicolumn{2}{c||}{\textbf{Number of  Samples}}                                                                                                                                          \\ \cline{2-3} 
    \multicolumn{1}{||c||}{}                                        & \multicolumn{1}{c|}{\textbf{\begin{tabular}[c]{@{}c@{}}Sequential\\ Splitting\end{tabular}}} & \multicolumn{1}{c||}{\textbf{\begin{tabular}[c]{@{}c@{}}Adjusted\\ Splitting\end{tabular}}} \\ \hline
    \hline
    $D^{train}_{raw}$                                             & \multicolumn{1}{l|}{7,401 (65\%)}                                                            & 7,387 (64.87\%)                                                                            \\ \hline
    $D^{val}_{raw}$                                               & \multicolumn{1}{l|}{1,708 (15\%)}                                                            & 1,694 (14.88\%)                                                                            \\ \hline
    $D^{test}_{raw}$                                              & \multicolumn{1}{l|}{2278 (20\%)}                                                             & 2,306 (20.25\%)                                                                            \\ \hline
    $D^{total}_{raw}$                                             & \multicolumn{1}{l|}{11,387 (100\%)}                                                          & 11,387 (100\%)                                                                             \\ \hline \hline
    $D^{train}_{dev}$                                             & \multicolumn{1}{l|}{7,231 (65.44\%)}                                                         & 7,093 (67\%)                                                                               \\ \hline
    $D^{val}_{dev}$                                               & \multicolumn{1}{l|}{1,609 (14.56\%)}                                                         & 1,444 (13.64\%)                                                                            \\ \hline
    $D^{test}_{raw}$                                              & \multicolumn{1}{l|}{2,209 (20\%)}                                                            & 2,050 (19.36\%)                                                                            \\ \hline
    $D^{total}_{dev}$                                             & \multicolumn{1}{l|}{11,049 (100\%)}                                                          & 10,587 (100\%)                                                                             \\ \hline \hline
    \end{tabular}
    }
\end{table}

\section{Evaluation Result}
This section evaluates the proposed GPS-assisted beam prediction 
and tracking method using the following key metrics:

\begin{enumerate}
    \item Top-K accuracy \cite{yeo-2024}: This metric quantifies the
    the proportion of time steps across test data set samples
    where the beam containing Top-K confidence scores encompasses
    the optimal beam. The Top-K accuracy is calculated
    as: 
    \begin{equation}
    \text{Top-}K \text{ accuracy} = \frac{1}{N^{test}}\sum_{l=1}^{N^{test}}\sum_{k=1}^{K} 1_{\{\hat{b}_{l,k}=b^*_l\}}.
    \label{eq:acc}
    \end{equation}
    Here, $N^{test}$ represents
    the total number of test samples, $\hat{b}_{l,k}$ is
    the $k_{th}$ predicted beam index for the $l$-th
    data sample, $b^*_l$ is the actual optimal beam
    index for the $l$-th sample, and $1_{\{\cdot\}}$ is
    an indicator function that returns one if the
    condition $\{\cdot\}$ is met.

    \item Average Power Loss \cite{morais-2023}: This metric assesses
    the average power loss between predicted and ground
    truth beams, mathematically expressed as: 
    \begin{equation}
    P_{L[dB]} = 10\log_{10}\left(\frac{1}{K}\sum_{k=1}^{K}\frac{P_{\mathbf{f}^*}^k-0.5P_n}{P_\mathbf{\hat{f}}^k-0.5P_n}\right), 
    \label{eq:powerloss}
    \end{equation}
    where $P_{n}$
    represents the noise power of the scenario, $P^k_{\mathbf{f}^*}$
    is the power of the ground truth beam
    in sample $k$, and $P^k_{\hat{\mathbf{f}}}$ is the power
    of the predicted beam for sample $k$. The noise power is 
    defined as the smallest power in sample $k$. The noise power $P_{n}$ is 
    multiplied by half to prevent zero 
    division in case of the predicted power $P^k_{\hat{\mathbf{f}}}$ has 
    equal value with noise power $P_{n}$.

    \item Overhead Savings \cite{morais-2023}: 
    This metrics is defined as the reduction in training beams, 
    are intrinsically linked to the desired reliability of the DL task. 
    Reliability is the degree of confidence we have that a particular 
    group of beams includes the optimal one. Higher reliability 
    necessitates a larger beam set and consequently diminishes 
    overhead savings. 
    The metric $\eta _{OH}^{(90)}$, representing the
    overhead savings for a 90\% reliability target, is
    calculated as $\eta_{OH}^{(90)}=1-\frac{b^{(90)}}{M}$, where $b^{(90)}$ denotes the
    minimum beams required to achieve this reliability and
    $M$ is the total codebook size.

    \item Average Power Loss Reliability: 
    The reliability of maintaining average power loss $P_{L[dB]}$
    below a threshold $P_{L[dB],treshold}$ is expressed 
    as $R(P_{L[dB],treshold})=\mathbb{P}(P_{L[dB]}\leq
    P_{L[dB],treshold})$, providing a probabilistic measure of the model's
    adherence to acceptable power loss margins.

    \item Model Size: This metric is calculated by multiplying the 
    number of non-zero model parameters ($N_{params}$) with a data 
    size ($Y$), which is set to $Y=32$ bits per model parameter.
\end{enumerate}

\subsection{Data Set Splitting Impact on Model Performance}

\begin{figure}[t]
    \centering
    \subfloat[]{%
        \includegraphics[clip,width=0.95\columnwidth]{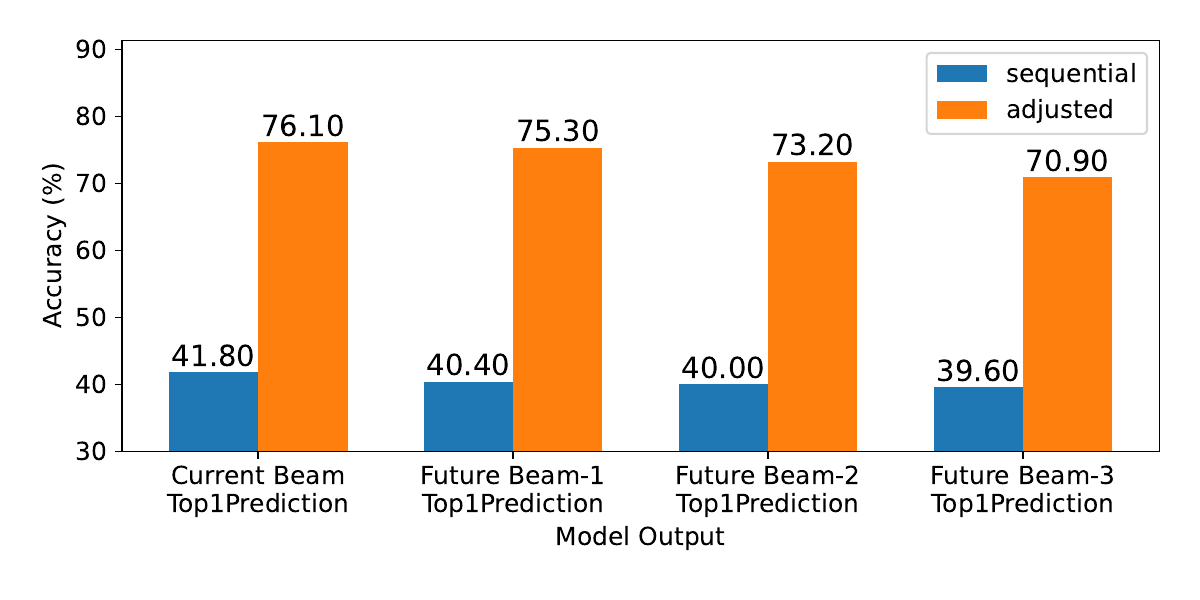}%
        \label{fig:acc-split}
    }\\[-0.8ex]
    \subfloat[]{%
        \includegraphics[clip,width=0.95\columnwidth]{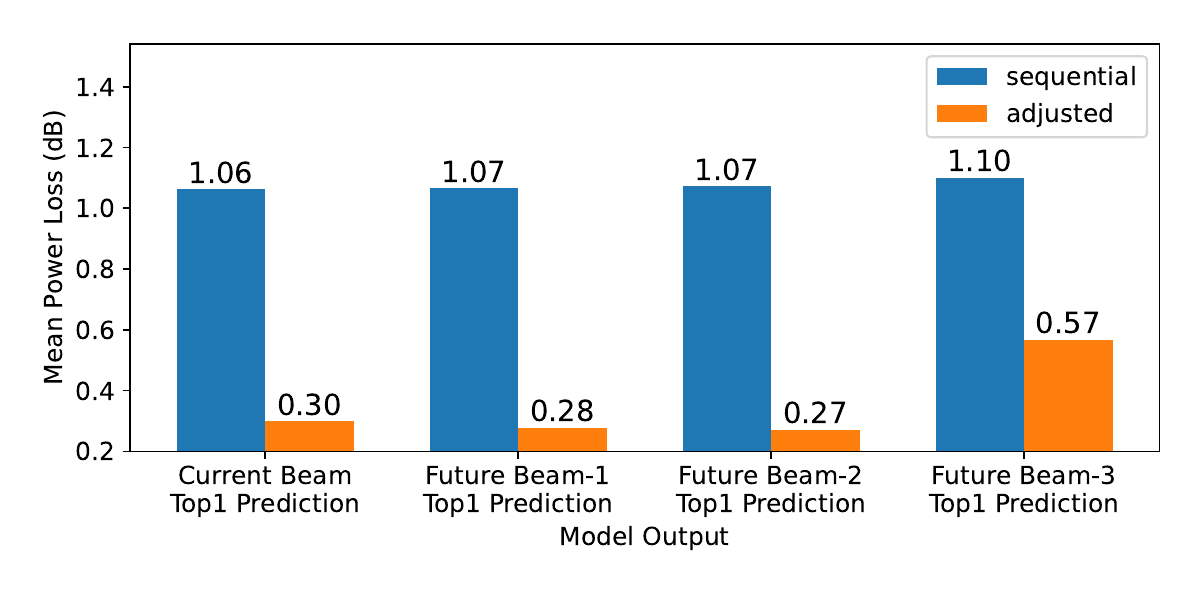}%
        \label{fig:power-split}
    }
    \caption{Impact of data set splitting methods on (a) model accuracy and (b) model mean power loss}
    \label{fig:result-split}
\end{figure}

In this experiment, we use concatenated normalized UE's geodetic position $\mathbf{g}_{UE,norm}$ and
UE-BS unit vector $\mathbf{u}_{UE-BS}$ as the input model but different data
set splitting method. The data set details are shown in Table \ref{tab:dataset-detail}.
The results in Figure \ref{fig:result-split}
reveal an interesting pattern regarding the impact of
data set splitting methods on accuracy and power
loss. For all model outputs, the adjusted splitting
method achieves higher accuracy and lower power loss
compared to the sequential method. The model accuracy
with the adjusted splitting is about $31\%\sim34\%$ higher
than with sequential splitting for current and future
beam predictions. Additionally, it achieves a power loss
reduction of around $48\%\sim70\%$ highlighting its effectiveness in
enhancing model performance. These findings emphasize the importance
of creating split data sets that have similar
label distribution with the original data set and
matches the train-validation-test split ratio of the data
set to enhance model performance.

\subsection{Model Performance Comparison}

\begin{table*}[]
    \caption{Comparison of model performance metrics across different model architecture and input}
    \centering
    \label{tab:model-comparison}
    \resizebox{0.95\textwidth}{!}{
    \begin{tabular}{||l||l||l||lll||lll||lll||lll||}
        \hline \hline
        \multirow{2}{*}{\textbf{\large{Model}}}                                    & \multirow{2}{*}{\textbf{\large{Input}}}                                                                      & \multirow{2}{*}{\begin{tabular}[c]{@{}l@{}}\textbf{\large{Size}}\\ \textbf{\large{(MB)}}\end{tabular}} & \multicolumn{3}{c||}{\textbf{Current Beam Prediction}}                                                                                                                                                                                                                   & \multicolumn{3}{c||}{\textbf{Future Beam-1 Prediction}}                                                                                                                                                                                                                  & \multicolumn{3}{c||}{\textbf{Future Beam-2 Prediction}}                                                                                                                                                                                                                  & \multicolumn{3}{c||}{\textbf{Future Beam-3 Prediction}}                                                                                                                                                                                                                  \\ \cline{4-15} 
                                                                       &                                                                                             &                                                                      & \multicolumn{1}{c|}{\begin{tabular}[c]{@{}c@{}}\textbf{Top-1}\\ \textbf{Acc.} \\ \textbf{(\%)}\end{tabular}} & \multicolumn{1}{c|}{\begin{tabular}[c]{@{}c@{}}\textbf{Top-3}\\ \textbf{Acc.} \\ \textbf{(\%)}\end{tabular}} & \multicolumn{1}{c||}{\begin{tabular}[c]{@{}c@{}}\textbf{Top-1 Mean} \\ \textbf{Power Loss}\\ \textbf{(dB)}\end{tabular}} & \multicolumn{1}{c|}{\begin{tabular}[c]{@{}c@{}}\textbf{Top-1}\\ \textbf{Acc.} \\ \textbf{(\%)}\end{tabular}} & \multicolumn{1}{c|}{\begin{tabular}[c]{@{}c@{}}\textbf{Top-3}\\ \textbf{Acc.} \\ \textbf{(\%)}\end{tabular}} & \multicolumn{1}{c||}{\begin{tabular}[c]{@{}c@{}}\textbf{Top-1 Mean} \\ \textbf{Power Loss}\\ \textbf{(dB)}\end{tabular}} & \multicolumn{1}{c|}{\begin{tabular}[c]{@{}c@{}}\textbf{Top-1}\\ \textbf{Acc.} \\ \textbf{(\%)}\end{tabular}} & \multicolumn{1}{c|}{\begin{tabular}[c]{@{}c@{}}\textbf{Top-3}\\ \textbf{Acc.} \\ \textbf{(\%)}\end{tabular}} & \multicolumn{1}{c||}{\begin{tabular}[c]{@{}c@{}}\textbf{Top-1 Mean} \\ \textbf{Power Loss}\\ \textbf{(dB)}\end{tabular}} & \multicolumn{1}{c|}{\begin{tabular}[c]{@{}c@{}}\textbf{Top-1}\\ \textbf{Acc.} \\ \textbf{(\%)}\end{tabular}} & \multicolumn{1}{c|}{\begin{tabular}[c]{@{}c@{}}\textbf{Top-3}\\ \textbf{Acc.} \\ \textbf{(\%)}\end{tabular}} & \multicolumn{1}{c||}{\begin{tabular}[c]{@{}c@{}}\textbf{Top-1 Mean} \\ \textbf{Power Loss}\\ \textbf{(dB)}\end{tabular}} \\ \hline \hline
        \begin{tabular}[c]{@{}l@{}}\large{Baseline} \\\large{Beam} \\ \large{Prediction}\end{tabular} & \begin{tabular}[c]{@{}l@{}}\large{$\mathbf{g}_{UE,norm}$}\end{tabular}                           & \large{1.07}                                                                 & \multicolumn{1}{l|}{\large{60}}                                                        & \multicolumn{1}{l|}{\large{87.5}}                                                      & \large{0.822}                                                                                        & \multicolumn{1}{c|}{-}                                                         & \multicolumn{1}{c|}{-}                                                         & -                                                                                            & \multicolumn{1}{c|}{-}                                                         & \multicolumn{1}{c|}{-}                                                         & -                                                                                            & \multicolumn{1}{c|}{-}                                                         & \multicolumn{1}{c|}{-}                                                         & -                                                                                            \\ \hline
        \begin{tabular}[c]{@{}l@{}}\large{Baseline} \\\large{Beam} \\ \large{Tracking}\end{tabular}   & \begin{tabular}[c]{@{}l@{}}\large{$\mathbf{g}_{UE,norm}$}\end{tabular}                           & \large{0.59}                                                                 & \multicolumn{1}{c|}{-}                                                         & \multicolumn{1}{c|}{-}                                                         & -                                                                                            & \multicolumn{1}{l|}{\large{63.9}}                                                      & \multicolumn{1}{l|}{\large{88.7}}                                                      & \large{0.905}                                                                                        & \multicolumn{1}{l|}{\large{63.9}}                                                      & \multicolumn{1}{l|}{\large{89.9}}                                                      & \large{1.068}                                                                                        & \multicolumn{1}{l|}{\large{63.2}}                                                      & \multicolumn{1}{l|}{\large{89.6}}                                                      & \large{1.348}                                                                                        \\ \hline
        \begin{tabular}[c]{@{}l@{}}\large{Proposed} \\ \large{Beam} \\ \large{Prediction} \\ \large{\&} \large{Tracking}\end{tabular}                                                       & \begin{tabular}[c]{@{}l@{}}\large{$\mathbf{g}_{UE,norm}$}\end{tabular}                           & \large{0.99}                                                                 & \multicolumn{1}{l|}{\large{64.8}}                                                      & \multicolumn{1}{l|}{\large{89.8}}                                                      & \large{0.726}                                                                                        & \multicolumn{1}{l|}{\large{63.7}}                                                      & \multicolumn{1}{l|}{\large{90.3}}                                                      & \large{0.805}                                                                                        & \multicolumn{1}{l|}{\large{63.4}}                                                      & \multicolumn{1}{l|}{\large{89.7}}                                                      & \large{0.889}                                                                                        & \multicolumn{1}{l|}{\large{62.5}}                                                      & \multicolumn{1}{l|}{\large{88.2}}                                                      & \large{1.224}                                                                                        \\ \hline
        \begin{tabular}[c]{@{}l@{}}\large{Proposed} \\ \large{Beam} \\ \large{Prediction} \\ \large{\&} \large{Tracking}\end{tabular}                                                       & \begin{tabular}[c]{@{}l@{}}\large{$\mathbf{u}_{UE-BS}$}\end{tabular}                                & \large{0.99}                                                                 & \multicolumn{1}{l|}{\large{62.1}}                                                      & \multicolumn{1}{l|}{\large{94.2}}                                                      & \large{0.953}                                                                                        & \multicolumn{1}{l|}{\large{61.3}}                                                      & \multicolumn{1}{l|}{\large{94.8}}                                                      & \large{0.862}                                                                                        & \multicolumn{1}{l|}{\large{59.3}}                                                      & \multicolumn{1}{l|}{\large{94.9}}                                                      & \large{0.819}                                                                                        & \multicolumn{1}{l|}{\large{57.5}}                                                      & \multicolumn{1}{l|}{\large{92.9}}                                                      & \large{1.077}                                                                                        \\ \hline
        \begin{tabular}[c]{@{}l@{}}\textbf{\large{Proposed}} \\ \textbf{\large{Beam}} \\ \textbf{\large{Prediction}} \\ \textbf{\large{\&} \large{Tracking}}\end{tabular}                                              & \begin{tabular}[c]{@{}l@{}}\{\large{$\mathbf{g}_{UE,norm}$,} \\ \large{$\mathbf{u}_{UE-BS}$}\}\end{tabular} & \large{\textbf{0.99}}                                                                 & \multicolumn{1}{l|}{\large{\textbf{76.1}}}                                    & \multicolumn{1}{l|}{\large{\textbf{97.3}}}                                             & \large{\textbf{0.299}}                                                                               & \multicolumn{1}{l|}{\large{\textbf{75.3}}}                                             & \multicolumn{1}{l|}{\large{\textbf{97.8}}}                                             & \large{\textbf{0.278}}                                                                               & \multicolumn{1}{l|}{\large{\textbf{73.2}}}                                             & \multicolumn{1}{l|}{\large{\textbf{97.4}}}                                             & \large{\textbf{0.272}}                                                                              & \multicolumn{1}{l|}{\large{\textbf{70.9}}}                                             & \multicolumn{1}{l|}{\large{\textbf{95.3}}}                                             & \large{\textbf{0.567}}                                                                                \\ \hline \hline
        \end{tabular}
    }
\end{table*}

The proposed beam prediction and tracking model (with
combined normalized UE's geodetic position $\mathbf{g}_{UE,norm}$ and UE-BS unit
vector $\mathbf{u}_{UE-BS}$ inputs) has a size of 0.99
MB while delivering meaningful performance. Performance 
metrics, detailed in
Table \ref{tab:model-comparison}, further highlight the model’s performance: Top-1
accuracy consistently exceeds 70\%, and Top-3 accuracy surpasses
95\% across all beam prediction steps, while the
mean power loss remains below 0.6 dB throughout.
This low power loss indicates minimal signal degradation,
even in cases where the optimal beam is
not selected.

We assess the proposed model alongside
baseline approaches from \cite{charan-2024} using the same data
set splitting method (adjusted data set splitting method)
and the same split ratio (65\% for training,
15\% for validation, and 20\% for testing). In
comparisons with these baseline methods, the proposed model
exhibits significant improvements in prediction accuracy and a
marked reduction in power loss, as presented in
Table \ref{tab:model-comparison}. Specifically, Top-1 accuracy increases by 7\%
$\sim$ 16\% across the four beam prediction steps,
with the greatest improvement about 16\% observed in
current beam prediction. Additionally, the Top-1 mean power
loss decreases by 57\% $\sim$ 69\% relative to
the baseline models.

Model input composition plays a
critical role in the model's performance improvements, as
shown in Table \ref{tab:model-comparison}. The combined approach using
both normalized UE's geodetic position $\mathbf{g}_{UE,norm}$ and UE-BS unit
vector $\mathbf{u}_{UE-BS}$ inputs outperforms single-input alternatives substantially. Compared
to using only normalized UE's geodetic position $\mathbf{g}_{UE,norm}$, the
dual-input approach increases Top-1 prediction accuracy by 8\%
$\sim$ 11.6\% across all prediction steps while decreasing mean power
loss by 53\% $\sim$ 69\%. Similarly, when compared
to using only UE-BS vector $\mathbf{u}_{UE-BS}$ input, the
combined approach yields 13\% $\sim$ 14\% improvements in
Top-1 accuracy and reduces mean power loss by
47\% $\sim$ 68\%. These results demonstrate that integrating
spatial information from both the UE position and
its directional relationship with the BS creates a
more comprehensive context for the model to make
accurate beam predictions.

\subsection{Model Performance Accross Various UAV Height}

\begin{figure*}[ht]
    \centering
    \includegraphics[width=0.9\textwidth]{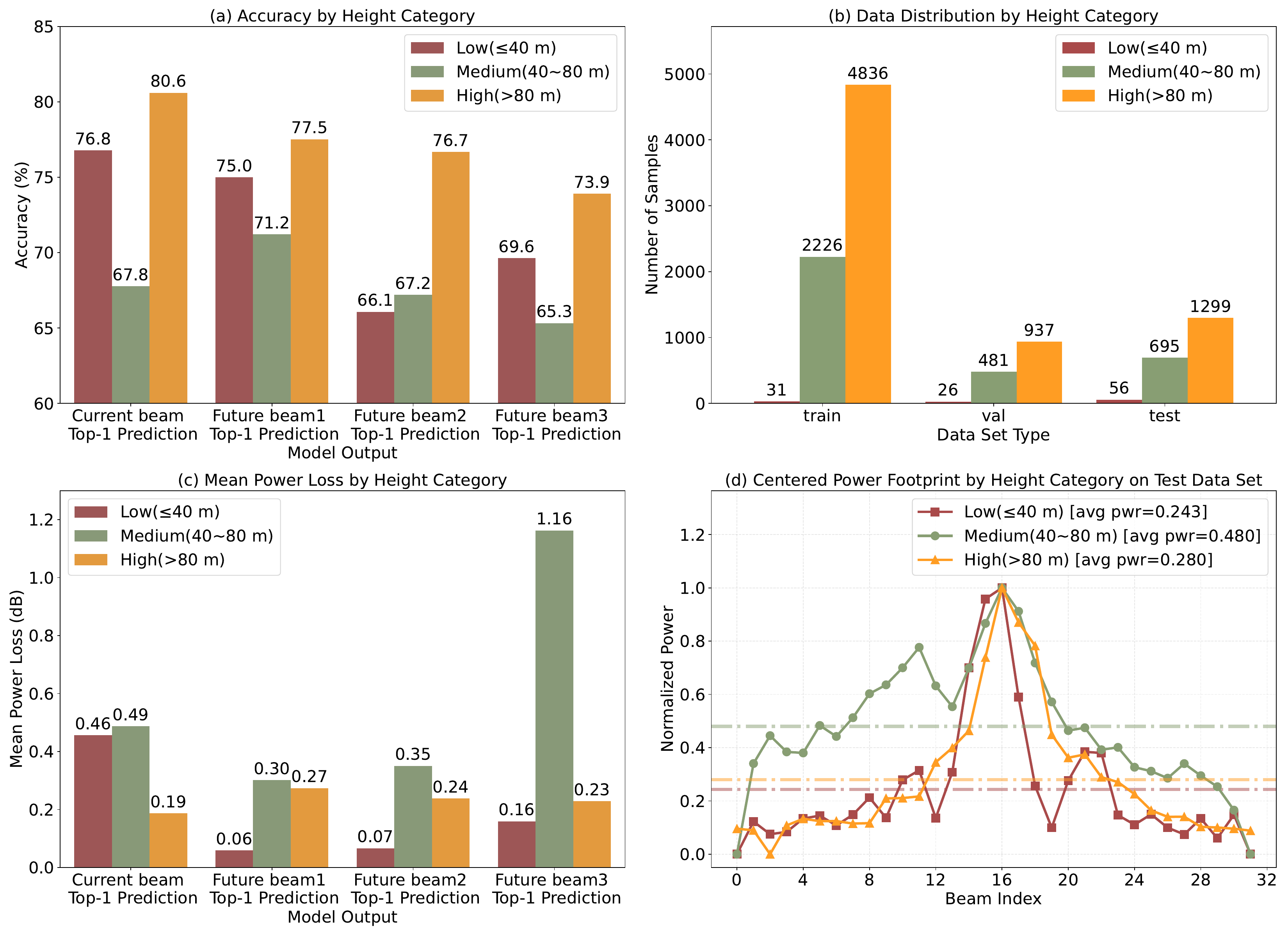}
    \caption{Model Performance on Various UAV Height}
    \label{fig:result-height}
\end{figure*}

Figure \ref{fig:result-height} (a) displays the accuracy of the
proposed model across different height categories, while Figure
\ref{fig:result-height} (b) illustrates the data set distribution across the
training, validation, and test sets. The model excels
in the high height category, with accuracy spanning
from about 73.9\% to 80.6\% across all beam
prediction steps, possibly reflecting its strong representation in
the data set (4,836 training samples). The low height
category, despite having the smallest training sample size
(31 samples), achieves a moderate accuracy range of
66.1\% to 76.8\% across all prediction steps. Meanwhile,
the medium height category demonstrates the lowest accuracy,
ranging from 65.3\% to 71.2\% across all steps.
To assess model performance independent of data set distribution,
we analyzed it using an equal sample size
(56 samples, sampled 20 times from the test
data set) across all height categories, as detailed in
Table \ref{tab:result-height}. The table indicates that the high
height category achieves the highest mean accuracy range
(71.88\% to 79.55\%), followed by the low (66.07\%
to 76.79\%) and medium categories (65.18\% to 73.39\%)
across all prediction steps. This trend implies that
accuracy does not increase linearly with flight height
and highlights medium heights as the most challenging
category.

Figure \ref{fig:result-height} (c) illustrates the mean power
loss across height categories for both current and
future beam predictions. For current beam predictions, the
medium height category exhibits the highest mean power
loss (0.49 dB), followed by low height category (0.46
dB) and high height category (0.19 dB). In future
beam predictions, the medium height category continues to
show the highest mean power loss range (0.3
dB to 1.16 dB), with high height category (0.23
dB to 0.27 dB) and low height category (0.06
dB to 0.16 dB) following. This suggests that
medium heights consistently suffer the largest mean power
loss across all prediction steps. This finding aligns
with the medium height category having the highest
average power value (0.480), as depicted in Figure
\ref{fig:result-height} (d). A higher average power loss indicates
a wider spread of power across beam indices,
which elevates noise power. According to equation \ref{eq:powerloss},
an increase in noise power narrows the gap
between predicted power and noise power, resulting in
a higher mean power loss. Overall, the pattern
reveals that medium heights consistently face greater power
loss challenges, driven by increased noise due to
power distribution.

\subsection{Model performance across various UAV speeds}

\begin{table*}[ht]
    \centering
    \caption{Model Performance on Various UAV Height with Equal Number of Samples}
    \label{tab:result-height}
    \resizebox{0.68\textwidth}{!}{
        \begin{tabular}{||l||l||l||l||l||l||l||}
        \hline \hline
        \multicolumn{1}{||c||}{\begin{tabular}[c]{@{}c@{}}Height \\ Category\end{tabular}} & \multicolumn{1}{c||}{\begin{tabular}[c]{@{}c@{}}Number \\ of \\ Samples\end{tabular}} & \multicolumn{1}{c||}{\begin{tabular}[c]{@{}c@{}}Sampling\\ Rounds\end{tabular}} & \multicolumn{1}{c||}{\begin{tabular}[c]{@{}c@{}}Current \\ Prediction\\ Top1 \\ Acc (\%)\end{tabular}} & \multicolumn{1}{c||}{\begin{tabular}[c]{@{}c@{}}Future \\ Prediction 1\\ Top1 \\ Acc (\%)\end{tabular}} & \multicolumn{1}{c||}{\begin{tabular}[c]{@{}c@{}}Future \\ Prediction 2\\ Top1 \\ Acc (\%)\end{tabular}} & \multicolumn{1}{c||}{\begin{tabular}[c]{@{}c@{}}Future \\ Prediction 3\\ Top1 \\ Acc (\%)\end{tabular}} \\ \hline
        \hline
        Low($\leq$40 m)                                                            & 56                                                                                   & 20                                                                             & 76.79±0.00                                                                                            & 75.00±0.00                                                                                             & 66.07±0.00                                                                                             & 69.64±0.00                                                                                             \\ \hline
        Medium(40$\sim$80 m)                                                             & 56                                                                                   & 20                                                                             & 69.29±6.10                                                                                            & 73.39±5.59                                                                                             & 69.38±4.53                                                                                             & 65.18±5.66                                                                                             \\ \hline
        High(\textgreater{}80 m)                                                         & 56                                                                                   & 20                                                                             & 79.55±3.89                                                                                            & 76.52±4.46                                                                                             & 74.91±6.38                                                                                             & 71.88±7.18                                                                                             \\ \hline \hline
        \end{tabular}
    }
\end{table*}

\begin{figure*}[t]
    \centering
    \includegraphics[width=0.9\textwidth]{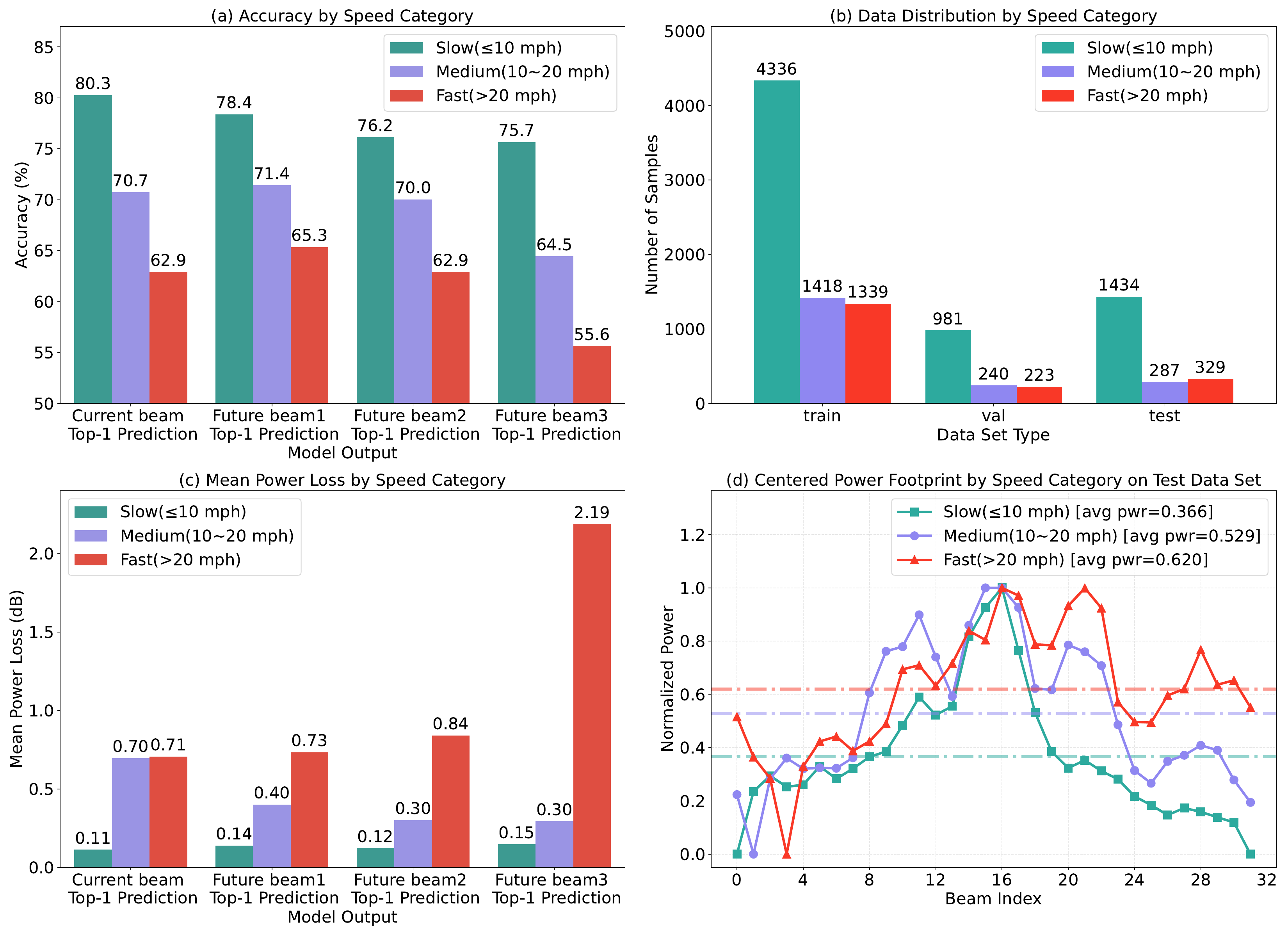}
    \caption{Model Performance on Various UAV Speed}
    \label{fig:result-speed}
\end{figure*}

Figure \ref{fig:result-speed} (a) illustrates the Top-1 prediction accuracy
of the proposed model across speed categories, while Figure \ref{fig:result-speed} (b) presents the
corresponding data set distribution across the training, validation, and
test sets. The model achieves the highest accuracy
in the slow speed category, ranging from approximately
75.7\% to 80.3\% for all beam prediction steps.
This may align with its dominant representation in
the data set (4,336 samples in the training set).
The medium speed category shows a lower accuracy
varying between 64.5\% and 71.4\% across all prediction
steps. The fast speed category exhibits the lowest
accuracy, ranging from 55.6\% to 65.3\% across all
prediction steps, despite having similar sample sizes to
the medium category (1,418 and 1,339 in the
training set for medium and fast categories, respectively).
This trend suggests that while the larger data set
size for the slow category may contributes to
its higher accuracy, the prediction accuracy steadily decreases
as speed increases. Moreover, despite the fast category
having a similar amount of data as the
medium category, its accuracy drops significantly, indicating that
higher speeds introduce prediction challenges not solely attributable
to low data samples.

Figure \ref{fig:result-speed} (c) depicts
the mean power loss across speed categories for
current and future beam predictions, while Figure \ref{fig:result-speed}
(d) shows the centered power footprint on the
test data set, with average power values of 0.366
for slow, 0.529 for medium, and 0.620 for
fast speed categories. In the slow speed category,
the mean power loss is low, varying between
0.11 dB and 0.15 dB, and the power
footprint is more concentrated at the center with
little variation across beam indices, showing that the
beam prediction and tracking remains steady and performs
reliably. For the medium speed category, the mean
power loss increases to a range of 0.3
dB to 0.7 dB, with a wider footprint
that fluctuates more, suggesting some difficulty in keeping
performance stable. Meanwhile, the fast speed category experiences
the highest power loss, from 0.71 dB to
2.19 dB, and its footprint is the most
spread out, with noticeable beam peaks, which points
to significant challenges in accurately predicting and tracking
beams at higher speeds. The trend of higher
mean power loss paired with a more spread-out
power footprint highlights how speed affects both the
model’s performance and the shape of the power
distribution, with faster speeds resulting in larger power
losses and wider power footprint.

\begin{figure*}[t]
    \centering
    \includegraphics[width=\textwidth]{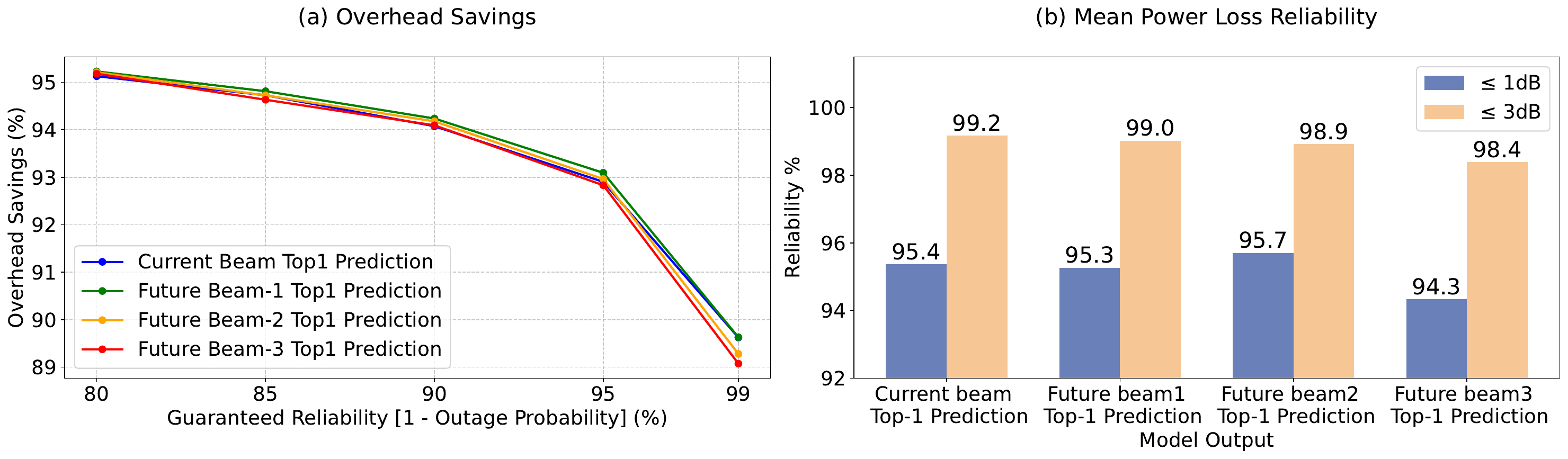}
    \caption{Model overhead savings and mean power loss reliability}
    \label{fig:overhead}
\end{figure*}

\begin{figure*}[ht]
    \centering
    \includegraphics[width=\textwidth]{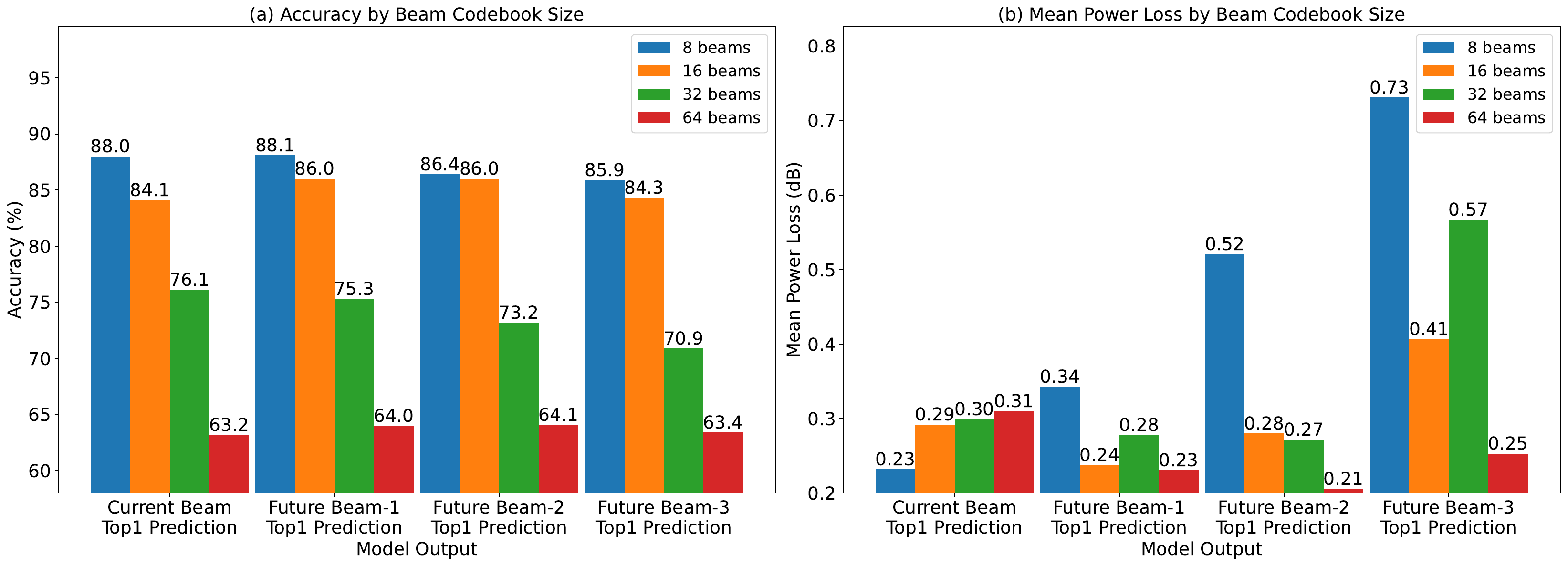}
    \caption{Model Performance on Various Beam Codebook Size}
    \label{fig:result-beam-codebook}
\end{figure*}

\subsection{Training Overhead Savings and Mean Power Loss Reliability }

Figure \ref{fig:overhead} (a) demonstrates how beam prediction and
tracking model reduces training overhead across different reliability
levels. At 80\% guaranteed reliability, the model achieves
approximately 95\% overhead savings, meaning that with a
codebook size of 32 classes, only 1 $\sim$
2 beams need training instead of all 32.
When the reliability threshold is elevated to 99\%,
the model still sustains a 89\% overhead savings,
necessitating the training of merely 3 $\sim$ 4
beams. This result highlight the model capability in
reducing beam training overhead. 

Figure \ref{fig:overhead} (b) analyzes
the mean power loss reliability across four beam
prediction steps, focusing on 1 dB and 3 dB thresholds.
When considering the stricter 1 dB threshold, which represents
a 20\% power loss, the reliability across all
prediction steps falls within a robust range of
94.3\% to 95.7\%. Moreover, when the threshold is
relaxed to 3 dB, signifying a 50\% power reduction,
the reliability across all prediction steps demonstrates higher
values, ranging from 98.4\% to 99.2\%. This result
highlights the model's capability in maintaining high power
loss reliability, particularly at less demanding thresholds.

\subsection{Model Performance Across Different Beam Codebook Sizes}

Figure \ref{fig:result-beam-codebook} evaluates model performance across beam codebook
sizes (8, 16, 32, and 64 beams). Figure
\ref{fig:result-beam-codebook}(a) reveals that smaller codebooks deliver higher beam
prediction accuracy, with 8 beams achieving a range
of 85\% $\sim$ 88\% and 16 beams ranging
from 84\% to 86\% for both current and
future beam predictions, with the current beam prediction
using 8 beams achieving the peak accuracy of
88.00\%. However, accuracy declines notably with larger codebooks,
starting with 32 beams where accuracy drops to
a range of 70\% $\sim$ 76\% across current
and future predictions, and further decreasing to a
range of 63\% $\sim$ 64\% for 64 beams.
The trend indicates that smaller beam codebook sizes
consistently outperform larger ones in terms of prediction
accuracy.

Figure \ref{fig:result-beam-codebook}(b) illustrates the mean power 
loss for both current and future beam predictions across varying codebook sizes. 
For current beam predictions, the mean power loss exhibits an upward trend, climbing 
from 0.23 dB in the 8-beam model to 0.31 dB in the 64-beam model. In the 8-beam 
model specifically, the mean power loss steadily increases with each prediction step, 
peaking at 0.73 dB for the third future beam prediction. The data reveals that while 
mean power loss in current beam predictions consistently grows with larger 
codebook sizes, future beam predictions show greater variability. Nevertheless, 
across all prediction steps, the mean power loss remains under 0.8 dB.

\section{Conclusion}
This study introduces a GPS-aided DL model 
capable of simultaneous prediction of both present and 
future beams for mmWave UAV communication, 
demonstrating performance with Top-1 accuracy exceeding 70\% 
and mean power loss below 0.6 dB throughout all temporal prediction steps.
The results are driven by a data set splitting strategy 
that ensures even label distribution, 
combined with a GPS preprocessing method that 
captures essential positional features, and a DL 
architecture that converts sequential position data 
into beam index predictions. The adjusted splitting 
approach outperforms sequential methods by 31\% $\sim$ 34\%
in accuracy while reducing mean power loss by
48\% $\sim$ 70\%. The combined feature approach—integrating 
normalized UE geodetic
position with UE-BS unit vector—consistently delivers superior performance,
improving Top-1 Accuracy by 8\% $\sim$ 11.6\% and decreasing mean
power loss by 53\% $\sim$ 69\% compared to using only
normalized UE geodetic position. Similar advantages (13\% $\sim$ 14\% higher
accuracy, 47\% $\sim$ 68\% lower mean power loss) are observed
over the model using only UE-BS unit vector input. Furthermore, the proposed model 
achieves 93\% overhead savings (requiring the training of 2 $\sim$ 3 beams 
instead of 32 beams)
while 
achieving beam prediction accuracy guarantees of 95\%. The 
model demonstrates reliable performance with 94\% $\sim$ 96\% 
of predictions maintaining power loss within 1 dB. 
Future research could aim to enhance the performance 
of the GPS-based model for UAVs operating at 
medium altitudes (40 m $\sim$ 80 m) and high speeds (>20 mph).

{\footnotesize
\bibliography{beamform-gps}}

\begin{thebibliography}{10}
\providecommand{\url}[1]{#1}
\csname url@samestyle\endcsname
\providecommand{\newblock}{\relax}
\providecommand{\bibinfo}[2]{#2}
\providecommand{\BIBentrySTDinterwordspacing}{\spaceskip=0pt\relax}
\providecommand{\BIBentryALTinterwordstretchfactor}{4}
\providecommand{\BIBentryALTinterwordspacing}{\spaceskip=\fontdimen2\font plus
\BIBentryALTinterwordstretchfactor\fontdimen3\font minus \fontdimen4\font\relax}
\providecommand{\BIBforeignlanguage}[2]{{%
\expandafter\ifx\csname l@#1\endcsname\relax
\typeout{** WARNING: IEEEtran.bst: No hyphenation pattern has been}%
\typeout{** loaded for the language `#1'. Using the pattern for}%
\typeout{** the default language instead.}%
\else
\language=\csname l@#1\endcsname
\fi
#2}}
\providecommand{\BIBdecl}{\relax}
\BIBdecl

\bibitem{zeng-2019}
Y.~Zeng, Q.~Wu, and R.~Zhang, ``Accessing {{From}} the {{Sky}}: {{A Tutorial}} on {{UAV Communications}} for {{5G}} and {{Beyond}},'' \emph{Proceedings of the IEEE}, vol. 107, no.~12, pp. 2327--2375, Dec. 2019.

\bibitem{nugroho-2025}
V.~A. Nugroho and B.~M. Lee, ``Gps-aided deep learning for beam prediction and tracking in uav mmwave communication,'' \emph{IEEE Access}, vol.~13, pp. 117\,065--117\,077, 2025.

\bibitem{xiao-2022}
Z.~Xiao, L.~Zhu, Y.~Liu, P.~Yi, R.~Zhang, X.-G. Xia, and R.~Schober, ``A {{Survey}} on {{Millimeter-Wave Beamforming Enabled UAV Communications}} and {{Networking}},'' \emph{IEEE Communications Surveys \& Tutorials}, vol.~24, no.~1, pp. 557--610, 2022.

\bibitem{geraci-2022}
G.~Geraci, A.~{Garcia-Rodriguez}, M.~M. Azari, A.~Lozano, M.~Mezzavilla, S.~Chatzinotas, Y.~Chen, S.~Rangan, and M.~D. Renzo, ``What {{Will}} the {{Future}} of {{UAV Cellular Communications Be}}? {{A Flight From 5G}} to {{6G}},'' \emph{IEEE Communications Surveys \& Tutorials}, vol.~24, no.~3, pp. 1304--1335, 2022.

\bibitem{xue-2024b}
Q.~Xue, C.~Ji, S.~Ma, J.~Guo, Y.~Xu, Q.~Chen, and W.~Zhang, ``A {{Survey}} of {{Beam Management}} for {{mmWave}} and {{THz Communications Towards 6G}},'' \emph{IEEE Communications Surveys \& Tutorials}, pp. 1--1, 2024.

\bibitem{zhang-2019a}
C.~Zhang, W.~Zhang, W.~Wang, L.~Yang, and W.~Zhang, ``Research {{Challenges}} and {{Opportunities}} of {{UAV Millimeter-Wave Communications}},'' \emph{IEEE Wireless Communications}, vol.~26, no.~1, pp. 58--62, Feb. 2019.

\bibitem{giordani-2019}
M.~Giordani, M.~Polese, A.~Roy, D.~Castor, and M.~Zorzi, ``A {{Tutorial}} on {{Beam Management}} for {{3GPP NR}} at {{mmWave Frequencies}},'' \emph{IEEE Communications Surveys \& Tutorials}, vol.~21, no.~1, pp. 173--196, 2019.

\bibitem{yi-2024}
W.~Yi, W.~Zhiqing, and F.~Zhiyong, ``Beam training and tracking in {{mmWave}} communication: {{A}} survey,'' \emph{China Communications}, vol.~21, no.~6, pp. 1--22, Jun. 2024.

\bibitem{chen-2019}
K.~Chen and C.~Qi, ``Beam {{Training Based}} on {{Dynamic Hierarchical Codebook}} for {{Millimeter Wave Massive MIMO}},'' \emph{IEEE Communications Letters}, vol.~23, no.~1, pp. 132--135, Jan. 2019.

\bibitem{qi-2020}
C.~Qi, K.~Chen, O.~A. Dobre, and G.~Y. Li, ``Hierarchical {{Codebook-Based Multiuser Beam Training}} for {{Millimeter Wave Massive MIMO}},'' \emph{IEEE Transactions on Wireless Communications}, vol.~19, no.~12, pp. 8142--8152, Dec. 2020.

\bibitem{alkhateeb-2014}
A.~Alkhateeb, O.~El~Ayach, G.~Leus, and R.~W. Heath, ``Channel {{Estimation}} and {{Hybrid Precoding}} for {{Millimeter Wave Cellular Systems}},'' \emph{IEEE Journal of Selected Topics in Signal Processing}, vol.~8, no.~5, pp. 831--846, Oct. 2014.

\bibitem{han-2016}
Y.~Han and J.~Lee, ``Two-stage compressed sensing for millimeter wave channel estimation,'' in \emph{2016 {{IEEE International Symposium}} on {{Information Theory}} ({{ISIT}})}, Jul. 2016, pp. 860--864.

\bibitem{alkhateeb-2018b}
A.~Alkhateeb, S.~Alex, P.~Varkey, Y.~Li, Q.~Qu, and D.~Tujkovic, ``Deep {{Learning Coordinated Beamforming}} for {{Highly-Mobile Millimeter Wave Systems}},'' \emph{IEEE Access}, vol.~6, pp. 37\,328--37\,348, 2018.

\bibitem{liu-2024a}
J.~Liu, X.~Li, T.~Fan, S.~Lv, and M.~Shi, ``Multimodal {{Fusion Assisted Mmwave Beam Training}} in {{Dual-Model Networks}},'' \emph{IEEE Transactions on Vehicular Technology}, vol.~73, no.~1, pp. 995--1011, Jan. 2024.

\bibitem{xin-2019}
X.~Xin and Y.~Yang, ``Robust {{Beam Tracking}} with {{Extended Kalman Filtering}} for {{Mobile Millimeter Wave Communications}},'' in \emph{2019 {{Computing}}, {{Communications}} and {{IoT Applications}} ({{ComComAp}})}, Oct. 2019, pp. 172--177.

\bibitem{larew-2019}
S.~G. Larew and D.~J. Love, ``Adaptive {{Beam Tracking With}} the {{Unscented Kalman Filter}} for {{Millimeter Wave Communication}},'' \emph{IEEE Signal Processing Letters}, vol.~26, no.~11, pp. 1658--1662, Nov. 2019.

\bibitem{ke-2019}
Y.~Ke, H.~Gao, W.~Xu, L.~Li, L.~Guo, and Z.~Feng, ``Position {{Prediction Based Fast Beam Tracking Scheme}} for {{Multi-User UAV-mmWave Communications}},'' in \emph{{{ICC}} 2019 - 2019 {{IEEE International Conference}} on {{Communications}} ({{ICC}})}, May 2019, pp. 1--7.

\bibitem{yuan-2020}
W.~Yuan, C.~Liu, F.~Liu, S.~Li, and D.~W.~K. Ng, ``Learning-{{Based Predictive Beamforming}} for {{UAV Communications With Jittering}},'' \emph{IEEE Wireless Communications Letters}, vol.~9, no.~11, pp. 1970--1974, Nov. 2020.

\bibitem{liu-2021c}
C.~Liu, W.~Yuan, Z.~Wei, X.~Liu, and D.~W.~K. Ng, ``Location-{{Aware Predictive Beamforming}} for {{UAV Communications}}: {{A Deep Learning Approach}},'' \emph{IEEE Wireless Communications Letters}, vol.~10, no.~3, pp. 668--672, Mar. 2021.

\bibitem{morais-2023}
J.~Morais, A.~Bchboodi, H.~Pezeshki, and A.~Alkhateeb, ``Position-{{Aided Beam Prediction}} in the {{Real World}}: {{How Useful GPS Locations Actually}} are?'' in \emph{{{ICC}} 2023 - {{IEEE International Conference}} on {{Communications}}}, May 2023, pp. 1824--1829.

\bibitem{jiang-2023}
S.~Jiang, G.~Charan, and A.~Alkhateeb, ``{{LiDAR Aided Future Beam Prediction}} in {{Real-World Millimeter Wave V2I Communications}},'' \emph{IEEE Wireless Communications Letters}, vol.~12, no.~2, pp. 212--216, Feb. 2023.

\bibitem{charan-2022a}
G.~Charan, T.~Osman, A.~Hredzak, N.~Thawdar, and A.~Alkhateeb, ``Vision-{{Position Multi-Modal Beam Prediction Using Real Millimeter Wave Datasets}},'' in \emph{2022 {{IEEE Wireless Communications}} and {{Networking Conference}} ({{WCNC}})}.\hskip 1em plus 0.5em minus 0.4em\relax Austin, TX, USA: IEEE, Apr. 2022, pp. 2727--2731.

\bibitem{luo-2023a}
H.~Luo, U.~Demirhan, and A.~Alkhateeb, ``Millimeter {{Wave V2V Beam Tracking}} using {{Radar}}: {{Algorithms}} and {{Real-World Demonstration}},'' in \emph{2023 31st {{European Signal Processing Conference}} ({{EUSIPCO}})}, Sep. 2023, pp. 740--744.

\bibitem{jiang-2022b}
S.~Jiang and A.~Alkhateeb, ``Computer {{Vision Aided Beam Tracking}} in {{A Real-World Millimeter Wave Deployment}},'' in \emph{2022 {{IEEE Globecom Workshops}} ({{GC Wkshps}})}, Dec. 2022, pp. 142--147.

\bibitem{charan-2022}
G.~Charan, A.~Hredzak, C.~Stoddard, B.~Berrey, M.~Seth, H.~Nunez, and A.~Alkhateeb, ``Towards {{Real-World 6G Drone Communication}}: {{Position}} and {{Camera Aided Beam Prediction}},'' in \emph{{{GLOBECOM}} 2022 - 2022 {{IEEE Global Communications Conference}}}, Dec. 2022, pp. 2951--2956.

\bibitem{charan-2023}
G.~Charan, A.~Hredzak, and A.~Alkhateeb, ``Millimeter {{Wave Drones}} with {{Cameras}}: {{Computer Vision Aided Wireless Beam Prediction}},'' in \emph{2023 {{IEEE International Conference}} on {{Communications Workshops}} ({{ICC Workshops}})}, May 2023, pp. 1896--1901.

\bibitem{ahmad-2023}
I.~Ahmad, A.~R. Khan, R.~N.~B. Rais, A.~Zoha, M.~A. Imran, and S.~Hussain, ``Vision-{{Assisted Beam Prediction}} for {{Real World 6G Drone Communication}},'' in \emph{2023 {{IEEE}} 34th {{Annual International Symposium}} on {{Personal}}, {{Indoor}} and {{Mobile Radio Communications}} ({{PIMRC}})}, Sep. 2023, pp. 1--7.

\bibitem{zarei-2023}
Z.~Zarei, F.~D. Tilahun, and C.~G. Kang, ``Vision-assisted {{Beam Prediction}} for {{UAV-enabled Millimeter-Wave Communications}} using {{SE-ResNet50}},'' in \emph{2023 14th {{International Conference}} on {{Information}} and {{Communication Technology Convergence}} ({{ICTC}})}, Oct. 2023, pp. 1659--1661.

\bibitem{charan-2024}
\BIBentryALTinterwordspacing
G.~Charan and A.~Alkhateeb, ``Sensing-{{Aided 6G Drone Communications}}: {{Real-World Datasets}} and {{Demonstration}},'' Dec. 2024. [Online]. Available: \url{http://arxiv.org/abs/2412.04734}
\BIBentrySTDinterwordspacing

\bibitem{alkhateeb-2023}
A.~Alkhateeb, G.~Charan, T.~Osman, A.~Hredzak, J.~Morais, U.~Demirhan, and N.~Srinivas, ``{{DeepSense 6G}}: {{A Large-Scale Real-World Multi-Modal Sensing}} and {{Communication Dataset}},'' \emph{IEEE Communications Magazine}, vol.~61, no.~9, pp. 122--128, Sep. 2023.

\bibitem{dong-2016}
\BIBentryALTinterwordspacing
W.~Dong, J.~Li, R.~Yao, C.~Li, T.~Yuan, and L.~Wang, ``Characterizing {{Driving Styles}} with {{Deep Learning}},'' Oct. 2016. [Online]. Available: \url{http://arxiv.org/abs/1607.03611}
\BIBentrySTDinterwordspacing

\bibitem{paul-1973}
M.~K. Paul, ``A note on computation of {{Geodetic}} coordinates from geocentric ({{Cartesian}}) coordinates,'' \emph{Bulletin g{\'e}od{\'e}sique}, vol. 108, no.~1, pp. 135--139, Jun. 1973.

\bibitem{zhu-1994}
J.~Zhu, ``Conversion of {{Earth-centered Earth-fixed}} coordinates to geodetic coordinates,'' \emph{IEEE Transactions on Aerospace and Electronic Systems}, vol.~30, no.~3, pp. 957--961, Jul. 1994.

\bibitem{ige-2024}
A.~O. Ige and M.~Sibiya, ``State-of-the-{{Art}} in {{1D Convolutional Neural Networks}}: {{A Survey}},'' \emph{IEEE Access}, vol.~12, pp. 144\,082--144\,105, 2024.

\bibitem{cho-2014}
K.~Cho, B.~Van~Merrienboer, C.~Gulcehre, D.~Bahdanau, F.~Bougares, H.~Schwenk, and Y.~Bengio, ``Learning {{Phrase Representations}} using {{RNN Encoder}}--{{Decoder}} for {{Statistical Machine Translation}},'' in \emph{Proceedings of the 2014 {{Conference}} on {{Empirical Methods}} in {{Natural Language Processing}} ({{EMNLP}})}.\hskip 1em plus 0.5em minus 0.4em\relax Doha, Qatar: Association for Computational Linguistics, 2014, pp. 1724--1734.

\bibitem{chung-2014}
\BIBentryALTinterwordspacing
J.~Chung, C.~Gulcehre, K.~Cho, and Y.~Bengio, ``Empirical {{Evaluation}} of {{Gated Recurrent Neural Networks}} on {{Sequence Modeling}},'' Dec. 2014. [Online]. Available: \url{http://arxiv.org/abs/1412.3555}
\BIBentrySTDinterwordspacing

\bibitem{yeo-2024}
Y.~Yeo and J.~Kim, ``Multi-{{Modal Sensing-Aided Beam Prediction}} using {{Poolformer}} for {{UAV Communications}},'' in \emph{2024 {{Fifteenth International Conference}} on {{Ubiquitous}} and {{Future Networks}} ({{ICUFN}})}, Jul. 2024, pp. 202--204.

\end{thebibliography}
\bibliographystyle{IEEEtran}%

\end{document}